\documentclass[prb,aps,amssymb,amsmath,reprint,superscriptaddress,showpacs]{revtex4-1}
\usepackage{bm}
\usepackage{graphicx}
\usepackage{color}

\newcommand{\beq}{\begin{equation}}
\newcommand{\eeq}{\end{equation}}
\newcommand{\beqa}{\begin{eqnarray}}
\newcommand{\eeqa}{\end{eqnarray}}

\newcommand{\w}{\omega}

\newcommand{\ket}[1]{\left| #1 \right\rangle}
\newcommand{\bra}[1]{\left\langle #1 \right|}

\newcommand{\akd}{a^{\dagger}_{k}}
\newcommand{\ak}{a^{\phantom{\dagger}}_{k}}

\begin{document}

\title{Unveiling environmental entanglement in strongly dissipative qubits
}

\author{Soumya Bera}
\author{Serge Florens}
\affiliation{Institut N\'{e}el, CNRS and UJF, B.P. 166, 25 Avenue des Martyrs,
38042 Grenoble, France}
\author{Harold U. Baranger}
\affiliation{Department of Physics, Duke University, Durham, North Carolina 27708, USA}
\author{Nicolas Roch}
\affiliation{Laboratoire Pierre Aigrain, \'{E}cole Normale Sup\'{e}rieure,
CNRS (UMR 8551), Universit\'{e} Pierre et Marie Curie,
Universit\'{e} Denis Diderot, 24 rue Lhomond, 75231 Paris Cedex 05, France}
\author{Ahsan Nazir}
\affiliation{Blackett Laboratory, Imperial College London, London SW7 2AZ, United Kingdom}
\author{Alex W. Chin}
\affiliation{Theory of Condensed Matter Group, University of Cambridge,
J J Thomson Avenue, Cambridge, CB3 0HE, United Kingdom} 

\begin{abstract}
The coupling of a qubit to a macroscopic reservoir plays a fundamental role in
understanding the complex transition from the quantum to the classical world.
Considering a harmonic environment, we use both intuitive arguments and
numerical many-body quantum tomography to study the structure of the complete
wavefunction arising in the strong-coupling regime, reached for intense
qubit-environment interaction.
The resulting strongly-correlated many-body ground state is built from quantum
superpositions of adiabatic (polaron-like) and non-adiabatic (antipolaron-like)
contributions from the bath of quantum oscillators. The emerging Schr\"odinger
cat environmental wavefunctions can be described quantitatively via simple
variational coherent states. In contrast to qubit-environment entanglement, we
show that non-classicality and entanglement among the modes in the reservoir are
crucial for the stabilization of qubit superpositions in regimes where standard
theories predict an effectively classical spin.
\end{abstract}

\date{\today}

\maketitle

The study of dissipative quantum phenomena, namely the interaction of a quantum
object (a qubit) with an infinite number of environmental degrees of freedom,
lies at the frontier of modern science and technology, with deep implications
for fundamental quantum physics~\cite{Raimond}, quantum
computing~\cite{nielsen}, and even
biology~\cite{LambertNP2013,scholes2011lessons}. While quantum information
stored in the qubit subsystem is lost during the coupling with the unobserved
degrees of freedom in the reservoir, it is in principle preserved in the entangled
many-body state of the global system.
The precise nature of this complete wavefunction has received little attention,
especially regarding the entanglement generated among the reservoir states.  Our
purpose here is to unveil a simple emerging structure of the wavefunctions in
open quantum systems, using a complementary combination of numerical many-body
quantum tomography and a novel analytical variational theory.

An archetype for quantitatively exploring the quantum dissipation 
problem~\cite{Leggett,Weiss,BreuerPetruccione} 
is to start with the simplest quantum object, a two-level system 
describing a generic quantum bit embodied by spin states 
$\left\{|\uparrow\rangle,|\downarrow\rangle\right\}$, 
and to couple it to an environment consisting of an infinite collection 
of quantum oscillators $\akd$
(with continuous quantum number $k$ and energy $\hbar \omega_k$). Quantum 
superposition of the two qubit states is achieved through a splitting
$\Delta$ acting on the transverse spin component, while dissipation (energy exchange 
with the bosonic environment) and decoherence 
are provided by a longitudinal interaction term $g_k$ with each displacement
field in the bath. 
This leads to the Hamiltonian of the celebrated continuum spin-boson model (SBM)~\cite{Leggett,Weiss}:
\begin{equation}
H = \frac{\Delta}{2} \sigma_x - 
\sigma_z \sum_k \frac{g_k}{2} (\akd+\ak) + \sum_k \w_k \akd \ak,
\label{ham}
\end{equation}
where we set $\hbar=1$, and the sums can be considered as integrals by introducing the 
spectral function of the environment, $J(\omega)\equiv\sum_{k}g_{k}^{2}\delta(\omega-\omega_{k})$. 
The generality of the SBM makes it a key model for
studying non-equilibrium dynamics, non-Markovian quantum evolution, 
biological energy transport, and the preparation and control of exotic quantum states in 
a diverse array of physical and chemical 
systems~\cite{scholes2011lessons,Leggett,Weiss,BreuerPetruccione,NitzanBook}.

The possibility of maintaining robust spin superpositions in the ground and 
steady states of the SBM has attracted considerable attention, 
primarily due to its 
implications for quantum computing \cite{Jennings09,RaussendorfBriegel2001}.
Previous numerical approaches have hitherto mainly focused on observables
related to the qubit degrees of freedom
\cite{NRG-RMP08,Makri95,WangThoss08,NalbachThorwart10,WinterBulla09,
AlvermannFehske09,PriorPlenio_EffSim10,Florens_DissipSpinDyn11}, 
whilst a description of the global system-environment wavefunction has been confined
to simpler variational studies \cite{Silbey,HarrisSilbey85,Chin_SubohmSBM11,Nazir,Demler}. 
This variational theory readily predicts the formation of semiclassical polaron states, 
which involve the adiabatic response of the environmental modes to the spin
tunneling.
Strong entanglement between the qubit and the bath is generated in this process.
We shall demonstrate here
that the many-body ground state of Hamiltonian~(\ref{ham}) contains additional 
non-classical correlations {\it among} the environmental oscillator
modes arising from their \emph{non-adiabatic} response to the spin-flip
processes. These new non-classical contributions to the wavefunction are key for 
the actual stabilization of qubit superpositions relative to the semiclassical picture, 
and naturally emerge from a variational framework beyond the adiabatic polaron approximation.



\begin{figure*}
\includegraphics[width=1.0\linewidth]{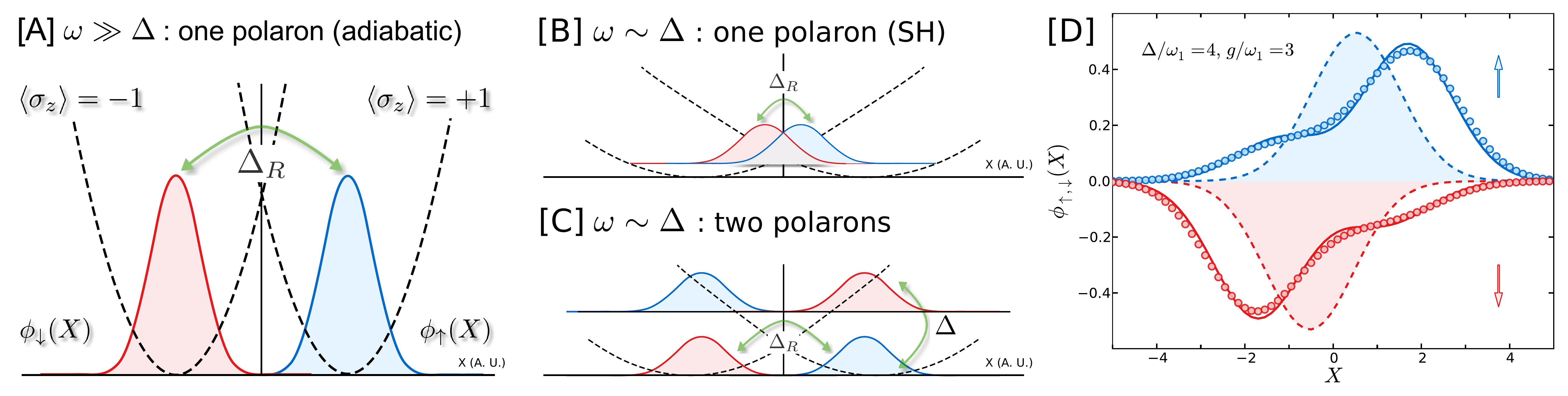}
\caption{{\bf Origins of polaron and antipolaron displacements in environment
wavefunctions.} In plots {\bf A-C}, black dashed lines are the spin-dependent
potential energies of a single harmonic oscillator in the absence of spin
tunneling [see Eq.\,(\ref{ham})], while blue (red) curves are the gaussian
wavefunctions (in real space $x$) of the oscillator on the $\langle
\sigma_x\rangle=1~(-1)$ potential surfaces. 
{\bf A}. {\bf Polarons}. For a high
frequency mode ($\omega \gg\Delta$), transitions to other oscillator states on
the potential surfaces are suppressed by the steep curvature of the potentials;
oscillator displacement adiabatically tunnels with the spin between minima of
the potentials, suppressing the tunneling amplitude by the reduced overlap of
the displaced oscillator wave functions to a value $\Delta_R$. 
{\bf B}. {\bf Non-adiabatic response in Silbey-Harris variational polaron theory}. Low
frequency modes ($\omega \ll\Delta$) have shallow potentials, leading to
well-separated minima. Poor wave function overlap prevents tunneling of the spin
between minima, destroying spin superposition. Variationally-determined
displacements adjust to smaller values, sacrificing their displacement
energy to maintain the spin-tunneling energy through better overlap. 
{\bf C}. {\bf Antipolaron response of non-adiabatic oscillators}. 
For modes with
$\omega \sim \Delta$, spin flips that do not change the position of the
oscillators (and thus have unsuppressed amplitude $ \Delta$) may become low
enough in energy to compete with the overlap-suppressed inter-minima tunneling.
The oscillator wavefunctions correlated with spin are now superpositions of
displaced coherent states with opposite signs. 
{\bf D}. {\bf Ground state wavefunction components of a single oscillator}. 
Spin up (blue) and spin down (red) components are shown for the exact ground
state (circles), our variational polaron-antipolaron state (solid lines), and
the Silbey-Harris ansatz (dashed lines) [$\Delta/\omega_1 =4$, $g/\omega_1 =3$]. 
The exact result shows distinct antipolaron features which are well captured by
the variational polaron-antipolaron state. The Silbey-Harris ansatz shows
reduced displacements and thus poor agreement with the exact result.}
\label{antipolaron}
\end{figure*}

In order to enlighten the nature of these emergent non-classical environmental states,
we first analyse the SBM by performing the (unitary) 
polaron transformation $\widetilde{H}= U H U^\dagger$, where $U=\exp\{-\sigma_z
\sum_k\frac{g_k}{2\w_k} (\akd-\ak)\}$, which removes the linear interaction 
term in Eq.~(\ref{ham}). This transforms the Hamiltonian to a basis 
in which oscillator wavefunctions are displaced according to the $z$-axis
projection of the spin:
\begin{equation}
\widetilde{H}= \frac{\Delta}{2} 
\sigma^+ e^{-\sum_k \frac{g_k}{\w_k} (\akd-\ak)} + {\rm h.c.}
+ \sum_k \w_k \akd \ak -E_{R},
\label{hamrot}
\end{equation}
where $E_{R}=\sum_{k}g_{k}^{2}/(4\omega_{k})$ is the reorganisation energy of
the bath. For $\Delta=0$, the ground state of $\widetilde{H}$ 
is doubly degenerate, and is given by the product of the bosonic vacuum and the spin states,
$|\widetilde{\Psi}_{\uparrow,0}\rangle= |\!\uparrow\rangle\otimes |0\rangle$ and $|\widetilde{\Psi}_{\downarrow,0}\rangle
=|\!\downarrow\rangle\otimes |0\rangle$, 
in the transformed basis (denoted by tildes). 
It thus corresponds to polaronic wavefunctions in the original frame,
where the positive/negative sign of the displacement is fully correlated to
the spin projection ({\it adiabatic} response):
$|\Psi_{\uparrow,g_k/2\omega_k}\rangle=
|\!\uparrow\rangle\otimes |+g_k/2\w_k\rangle$ and $|\Psi_{\downarrow,-g_k/2\omega_k}\rangle=
|\!\downarrow\rangle\otimes |-g_k/2\w_k\rangle$.
The two-fold degenerate ground state thus takes the
form of a product of semiclassical coherent states (displaced oscillators) 
$|\pm f_k\rangle \equiv e^{\pm\sum_k f_k (\akd-\ak)}|0\rangle$,
with displacements $f_k=\pm g_{k}/2\omega_{k}$ which shift each oscillator to the 
minimum of its static spin-dependent potential. 
This potential is evident in Eq.\,(\ref{ham}) for $\Delta=0$ and
is shown explicitly in Fig.~\ref{antipolaron}{\bf A}. In the presence of spin
tunneling ($\Delta\neq0$), one needs to understand the effect of the operators
$K_{\pm}\equiv \Delta \sigma^\pm e^{\mp\sum_k(g_k/\w_k) (\akd-\ak)}$ in
Eq.~(\ref{hamrot}) which
correlate spin flip processes with
simultaneous displacements of \emph{all} oscillator states. As we now show, 
these correlations
ultimately control the ground state qubit superposition.

\emph{Polarons, antipolarons, and ground state ansatz}. 
The optimum oscillator displacements result from a competition between 
the two terms appearing in Hamiltonian~(\ref{hamrot}), namely spin tunneling
$\Delta$ versus oscillator kinetic energy.
Within the transformed frame, consider the coupling induced by 
the tunneling operator $K_{\pm}$ between one of the doubly degenerate states, 
say $|\widetilde{\Psi}_{\downarrow,0}\rangle=|\downarrow\rangle\otimes |0\rangle$, 
and a spin-flipped state with arbitrary
displacement function $\widetilde{f}_k\equiv f_k-g_k/2\omega_k$, 
$|\widetilde{\Psi}_{\uparrow,\widetilde{f}_k}\rangle=| \uparrow\rangle\otimes| \widetilde{f}_{k}\rangle$
($f_k$ is the displacement in the original frame). 
The matrix element for this is 
\begin{equation}\label{renorm}
\langle \widetilde{\Psi}_{\uparrow,\widetilde{f}_k}|K_{+}|\widetilde{\Psi}_{\downarrow,0}\rangle=\Delta 
e^{- \frac{1}{2}\sum_{k}(\widetilde{f}_{k}+g_{k}/\omega_{k})^2}.
\end{equation}
The elastic displacement energy of oscillator $k$ in
$|\widetilde{\Psi}_{\uparrow,\widetilde{f}_k}\rangle$ is $\langle
\widetilde{\Psi}_{\uparrow,\widetilde{f}_k}|\omega_{k}a_{k}^{\dagger}a_{k}|
\widetilde{\Psi}_{\uparrow,\widetilde{f}_k}\rangle=\omega_{k}\widetilde{f}_{k}^{2}$.
The polaron transformed Hamiltonian reveals the inherent competition between the
(elastic) energetic cost of mixing displaced oscillators into the ground state, favouring 
$\widetilde{f}_{k}=0$, and the exponential suppression of the spin kinetic 
energy, given by the reduced tunneling matrix element in Eq.~(\ref{renorm}), which
rather favours $\widetilde{f}_{k}=-g_{k}/\omega_{k}$. For high-frequency modes,
the elastic energy cost dominates and tunneling between
spin states is governed by environment states with
$\widetilde{f}_{k}=0$, gaining only a small (renormalized) tunneling energy 
$\Delta_R=\Delta e^{- \frac{1}{2}\sum_{k}(g_{k}/\omega_{k})^2}\ll\Delta$
for strong qubit-environment interaction. In the 
original frame, the corresponding displacement is
$f_k=+g_k/2\omega_k$, which implies that these 
`fast' oscillators instantaneously (adiabatically) tunnel with the spin between the 
minima of their elastic potentials -- see Fig.~\ref{antipolaron}.
In the opposite limit of low-frequency modes ($\omega_{k}\ll\Delta_R$), the elastic 
energy barrier is weak; mixing between spin states is instead governed by the matrix 
element~(\ref{renorm}). Returning to the original frame, one gets an energy gain 
of the {\it bare} tunneling energy $\Delta$ when $f_{k}=-g_{k}/2\omega_{k}$. 
As shown in Fig.~\ref{antipolaron}{\bf C}, this corresponds to spin tunneling
with {\it non-adiabatic response} of the oscillators, which are 
displaced in the \emph{opposite} direction from the adiabatic modes.
We naturally dub these contributions to the wavefunction 
\emph{antipolaron states}. 
At intermediate frequencies, we expect that both polaronic and 
antipolaronic responses occur, leading to a
two-polaron ansatz for the ground state (in the original frame): 
%
\begin{eqnarray}
\label{trial}
\big|GS^{\mathrm{2pol.}}\big> &=& \big|\uparrow\big>
\otimes \left[ 
\big|\!+\!f^{\mathrm{pol.}}_{k}\big> +
p \big|\!+\!f^{\mathrm{anti.}}_{k}\big> \right]\\
\nonumber
&-& \big|\downarrow\big> 
\otimes \left[ 
\big|\!-\!f^{\mathrm{pol.}}_{k}\big> +
p  \big|\!-\!f^{\mathrm{anti.}}_{k}\big> \right],
\end{eqnarray}
with $p$ the relative weight of the polaron and antipolaron components.
Note that this ansatz fully respects the symmetries of the Hamiltonian. 

This state reduces to standard (adiabatic) polaron theory when $p=0$ and
$f^{\mathrm{pol.}}_{k}=g_{k}/2\omega_k$, and to the variational polaron state of Silbey and
Harris (SH) \cite{Silbey,HarrisSilbey85} when $p=0$ and the function $f^{\mathrm{pol.}}_{k}$ is 
varied to minimise the total ground state energy 
$E=\langle GS^{\mathrm{2pol.}}| H| GS^{\mathrm{2pol.}}\rangle$. 
As we shall compare our ansatz~(\ref{trial}) to these simpler theories, a brief 
description of them is given in the Supplementary Information. For $p\neq0$, 
the environment wave function for each spin projection is a
multi-modal Schr\"odinger cat state involving a superposition of polaronic and antipolaronic components, 
leading to considerable mode entanglement. 
The critical observation is that such superposition of displaced states lowers the 
energy of the ground state by \emph{stabilising} the spin energy. 
For the state (\ref{trial}) the spin tunneling energy 
$E_{T}=(\Delta/2) 
\langle GS^{\mathrm{2pol.}}| \sigma_x| GS^{\mathrm{2pol.}}\rangle$ is
\begin{eqnarray}
\label{TunnelEnergy}
E_T&=&-\Delta e^{-2\sum_{k} (f^{\mathrm{pol.}}_{k})^{2}}
-p^2\Delta e^{-2\sum_{k}(f^{\mathrm{anti.}}_{k})^{2}}\nonumber\\
&-&2p\Delta e^{-\frac{1}{2}\sum_{k}(f^{\mathrm{pol.}}_{k}+f^{\mathrm{anti.}}_{k})^{2}}.
\end{eqnarray}
The first two terms reflect an exponentially suppressed renormalized tunneling rate. 
Indeed, for strong coupling, the displacements $f^{\mathrm{pol.}}_{k}$ and $f^{\mathrm{anti.}}_{k}$ are large,
and the associated contribution to the spin energy becomes vanishingly small.
However, the overlap between the polaron and antipolaron contributions (the third
term) will not be suppressed if, as we expect, $f^{\mathrm{anti.}}_{k}\approx -f^{\mathrm{pol.}}_{k}$. The
development of a small but finite antipolaron weight $p$ thus allows the environment to 
minimise its displacement energy whilst maintaining significant overlap 
between the environment-dressed spin states. 

\emph{Single mode}. 
Before tackling the challenging many-mode situation, we develop intuition 
about the polaron-antipolaron ansatz in the simplest case of a single 
environmental mode with energy $\omega_1$ and coupling $g_1$. This case is easily
diagonalised numerically (see also Ref.~\onlinecite{Braak} for an exact solution); 
note that a similar ansatz for the single-mode Rabi model (without reference to polaron theory) 
has been previously explored numerically \cite{hwang,stolze}. 
In Figure~\ref{antipolaron}{\bf D} 
we compare the spatial wavefunctions of the oscillator
correlated with each spin state with those obtained from the ansatz Eq.\,(\ref{trial}) following a numerical 
optimization of
$p$, $f^{\mathrm{pol.}}_{1}$, and $f^{\mathrm{anti.}}_{1}$ to minimise
the ground state energy. Choosing oscillator parameters where we expect
non-adiabatic response, namely $\omega_1<\Delta$, we find that both 
wavefunctions clearly show
a superposition of polaron and antipolaron contributions, with much 
larger displacements compared to the prediction of the SH theory (single polaron 
case $p=0$). The agreement of the diagonalised and the two-polaron ansatz
ground state wavefunctions is extremely good, as well as the energies and spin
observables, even for a coupling strength as large as $g=3\omega_1$ (see also  
Supplementary Information).
As motivated above, the emergence of an antipolaron component in the environment
enhances the overlap of the tunneling states. 
The single polaron SH state fails in this regard (see
Figures~\ref{antipolaron}{\bf B} and 1{\bf D}, and Supplementary Information)
as it finds itself frustrated between 
minimizing the elastic energy and maintaining good overlap between the opposite 
spin states: the resulting displacements are thus totally wrong. 

\begin{figure*}[tbp]
\includegraphics[width=1.0\linewidth]{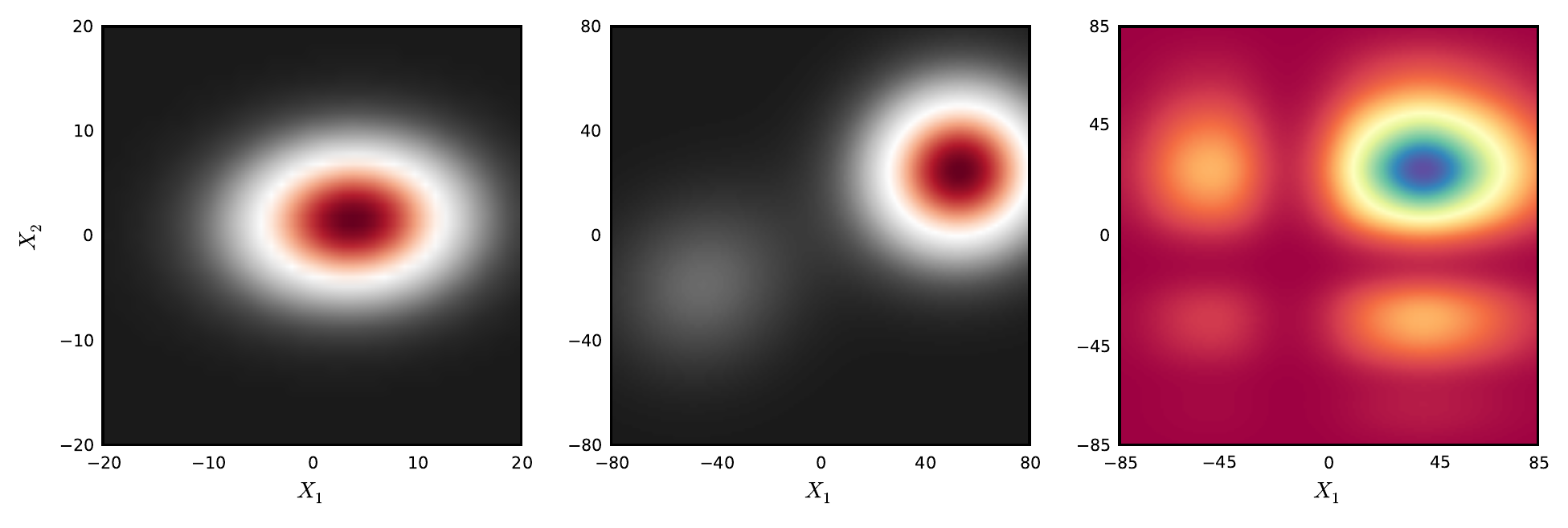}
\caption{{\bf Two mode wavefunctions}. {\bf A-B}. Contour plots in real space of 
the spin-up projected joint oscillator wavefunctions of two modes 
obtained from exact diagonalisation.
{\bf A. Polarons.} For high frequency modes 
($\omega_2 = 2\omega_1 = 0.04 > \Delta =0.01$), the wavefunction is a single, 
displaced gaussian, in qualitative agreement with Silbey-Harris theory.
{\bf B. Entangled antipolarons.} Low frequency modes 
($\omega_2 = 2\omega_1 = 0.004 < \Delta=0.01$) show the development 
of an antipolaron component, visible in the $(X_1<0,X_2<0)$ quadrant, in addition 
to the Silbey-Harris state.
{\bf C. Product state.} Hypothetical wavefunction obtained from a 
product state of polaron-antipolaron superpositions for each mode,
showing symmetric off-diagonal peaks. These features are absent in {\bf B}, 
indicating that the exact joint wavefunction is not a product state but, 
in contrast, is entangled as described in the text.}
\label{twomodes}
\end{figure*}

\emph{Two-mode antipolaron entanglement}. Having confirmed the emergence of
non-adiabatic antipolaron contributions in the case of a single mode, we now
consider the case of a two-mode SBM and, in particular, 
test our proposal Eq.~(\ref{trial}) that the two-mode wave function 
dressing a given spin state will be entangled.

Fig.~\ref{twomodes} shows the spin-up component of the two-mode wavefunction 
as a function of the two independent spatial coordinates of the modes 
($x_1$ and $x_2$) for two modes taken at different frequencies $\omega_2=2\omega_1$.
The ground state wavefunctions were determined by exact numerical diagonalisation.
We see the clear development of an antipolaron component to the wavefunction 
(Fig.~\ref{twomodes}{\bf B}) for low-energy non-adiabatic modes, in contrast to
the situation of high-energy adiabatic modes (Fig.~\ref{twomodes}{\bf A}).
However, we see that only two peaks appear in the wavefunction -- those along the diagonal line
$x_{1}=x_{2}$ -- indicating unambiguously that this two-mode wavefunction takes
the inter-mode entangled form 
$|f_1^{\mathrm{pol.}}\rangle\otimes|f_2^{\mathrm{pol.}}\rangle
+p|f_1^{\mathrm{anti.}}\rangle\otimes|f_2^{\mathrm{anti.}}\rangle$.
This can be contrasted with a hypothetical polaron-antipolaron product 
state
$\big\{|f_1^{\mathrm{pol.}}\rangle+p|f_1^{\mathrm{anti.}}\rangle\big\}\otimes
\big\{|f_2^{\mathrm{pol.}}\rangle+p|f_2^{\mathrm{anti.}}\rangle\big\}
$
which would rather display four peaks, as shown in Fig.~\ref{twomodes}{\bf C}. 
The implications of this inter-mode entanglement for the entropy of the reservoir 
modes is given in Supplementary Information. Again, one can check that the variational 
energy of the two-mode ground state 
is remarkably close to the exact energy. 

\emph{Multi-mode spin-boson model}. We now turn to the more challenging 
many-mode situation, tackling the continuum spin-boson model~(\ref{ham}).
A direct diagonalisation of the Hamiltonian is now hopeless; however, recent
computational progress has opened the way to calculating ground state averages
of arbitrary operators, for instance using the bosonic Numerical Renormalization Group
(NRG)~\cite{Bulla}, which we will use to test the generalized polaron state~(\ref{trial}). 
A key feature in the NRG method is the use of a logarithmic discretization of
the energy spectrum of the bath, which ensures the stability and convergence of an 
iterative diagonalization of the impurity model~\cite{NRG-RMP08}. In order to directly 
compare with the variational results, we use the same discretization in defining the polaronic 
ansatz Eq.~(\ref{trial}), incorporating the changing measure in $\omega_k$ into the
definitions of the $f_k$.

We focus here on the standard case of ohmic dissipation \cite{Leggett,Weiss},
although our following results should apply similarly to other types of spectral density.
The continuous bath of bosonic excitations assumes then a linear spectrum in frequency, 
$J(\w) = 2 \alpha \w \theta(\w_c-\w)$,
up to a high energy cutoff $\w_c$ and with dimensionless dissipation 
strength $\alpha$. 
Weakly damped Rabi oscillations of the qubit for $\alpha\ll1$ are known 
to completely fade away in the strong dissipation regime $\alpha \gtrsim 0.4$, 
where the qubit becomes strongly entangled with its environment. 
The bare qubit frequency $\Delta$ is heavily renormalized
in this regime to the smaller value $\Delta_R=\Delta (\Delta
e/\w_c)^{\alpha/(1-\alpha)}$, for $\Delta/\w_c\ll1$, which can thus be driven to 
zero for the critical dissipation strength $\alpha_c\simeq1$, indicating a 
quantum critical point.

\begin{figure*}[t]
\includegraphics[width=0.9\columnwidth]{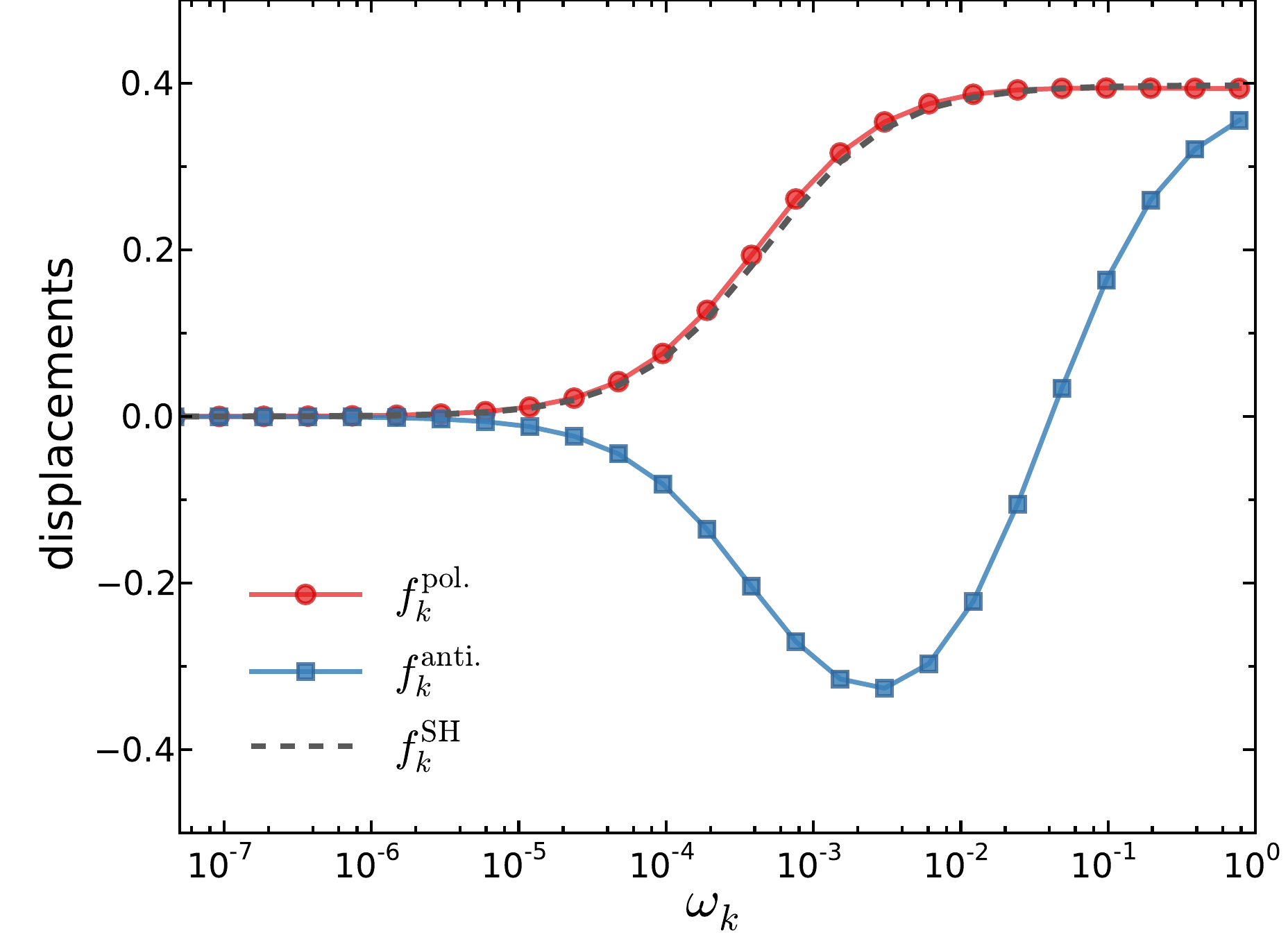}
\includegraphics[width=0.9\columnwidth]{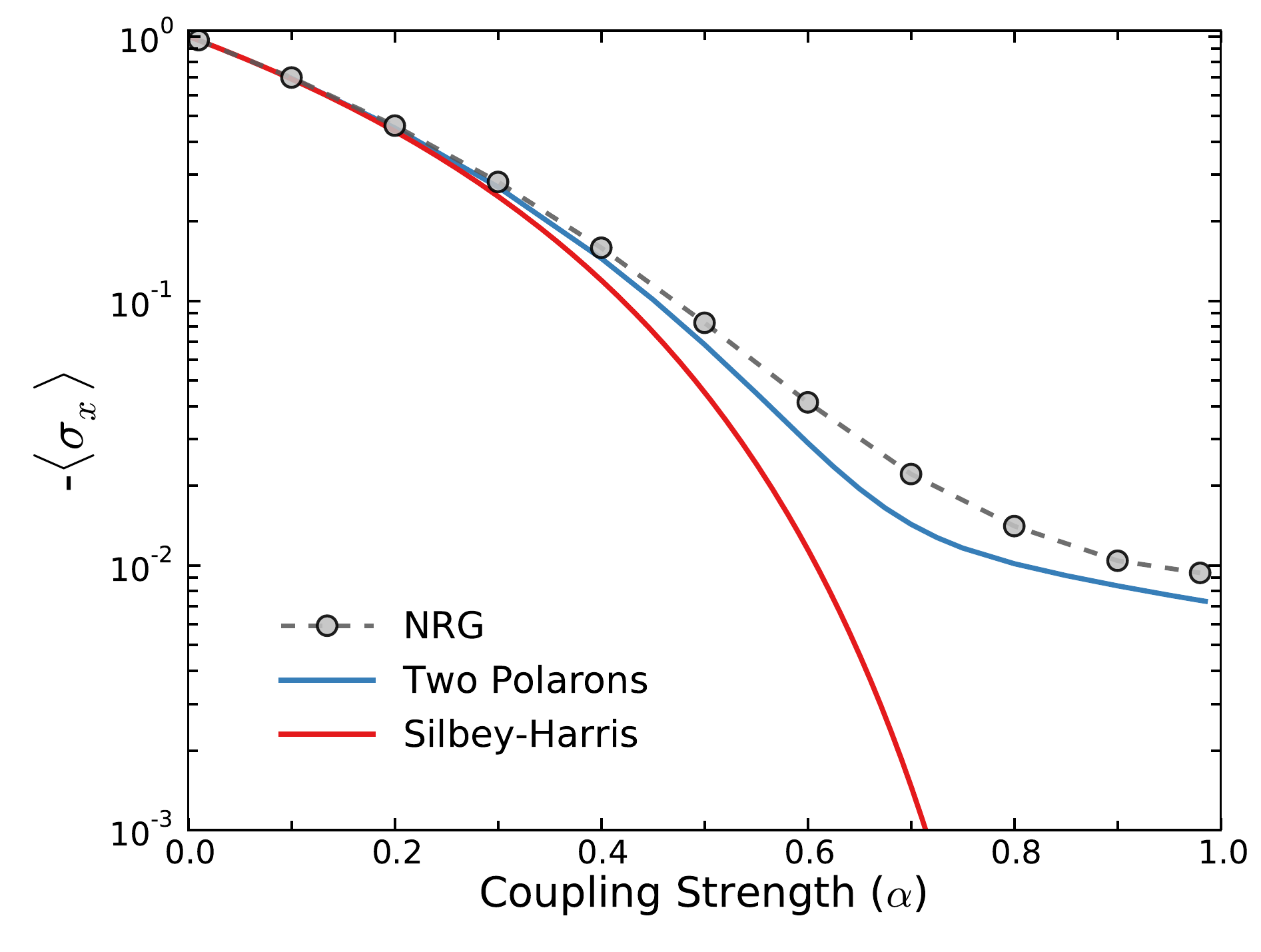}
\caption{{\bf Displacements and spin average in the many-mode case.} 
{\bf A.} Displacements determined variationally from the two-polaron ansatz 
Eq.~(\ref{trial}), showing the emergence of an antipolaron component for low 
energies, with equal and opposite displacement to the polaron state.
The antipolaron state merges smoothly onto the polaron state at high energy as the 
adiabaticity of the oscillators with respect to tunneling of the spin is
recovered
(the NRG logarithmic discretisation of the bath spectrum is used here, namely
frequency points are evenly spaced on a logarithmic scale; Note that a point at higher 
energy is associated to a larger energy window of the continuum spectrum, leading to 
the saturation of $f_k^\textrm{pol.}$ for high frequencies, instead of the fall off 
obtained for a linear energy mesh). 
[Parameters: $\alpha=0.5$ and $\Delta=0.01$.]
{\bf B.} Ground state averaged spin amplitude $-\big<\sigma_x\big>$ as a function
of dissipation strength $\alpha$ computed with the NRG (circles) for $\Delta/\w_c=0.01$, 
and compared to the one-polaron (red line) and two-polaron (blue line) 
predictions. 
A clear breakdown of the one-polaron Silbey-Harris ansatz occurs at strong dissipation, 
while the two-polaron trial state accounts for the correct behavior up to the 
quantum critical point ($\alpha_c=1$), due to preserved tunneling amplitude via 
the antipolaron component of the wavefunction.}
\label{ManyModesSx}
\end{figure*}

As a first step towards understanding the many-mode situation, we consider
the variational solution obtained from the two-polaron 
ansatz~(\ref{trial}) (the variational equations are given in the Supplementary Information).
This leads to the polaronic and antipolaronic displacements shown in
Fig.~\ref{ManyModesSx}{\bf A}, which exemplify the physical picture introduced
above (see especially Fig.~\ref{antipolaron}): polaron and antipolaron
states show equal and opposite displacements at low energies
(typically for $\omega_k\ll\omega_c$), but merge together to produce a fully
polaronic state at high energy, where the environment responds adiabatically
to the spin. The variational theory is thus able, without additional physical
input, to generate the correct crossover from non-adiabatic to adiabatic
behavior of the antipolaron component with increasing energy.

The presence of the antipolaron component has a large impact on the ground state
spin average: Fig.~\ref{ManyModesSx}{\bf B} compares the result of the one- and
two-polaron variational states to that computed with NRG (numerically exact
result, used as a benchmark). 
In the one-polaron (SH) limit ($p=0$), one finds readily
$ -\big<\sigma_x\big> = 
\Delta_R/\Delta = (\Delta e/\w_c)^{\alpha/(1-\alpha)}$, which incorrectly vanishes 
at the critical dissipation strength $\alpha_c=1$~\cite{Silbey,Nazir}.
%
On the other hand, the emergence of antipolaron correlations at low energy,
namely $f^{\mathrm{anti.}}_{k}\simeq -f^{\mathrm{pol.}}_{k}$ for 
$\w_k\ll \omega_c$, helps in maintaining a 
finite value for $\big<\sigma_x\big>$, due to the perfect cancellation of
the displacements within the exponential in the last term of
Eq.~(\ref{TunnelEnergy}).
This success of the antipolaron ansatz~(\ref{trial}) is illustrated in 
Fig.~\ref{ManyModesSx}{\bf B}.

Our objective now is to demonstrate the peculiar inter-mode entanglement 
properties of the antipolaron ansatz~(\ref{trial}).
While one cannot plot the complete many-body wavefunction in the case of 
many environmental modes, a useful strategy 
to assess the validity of the trial state~(\ref{trial}) lies in recent interest 
in quantum tomography~\cite{ReviewTomography,Hofheinz,Eichler}, wherein the
reduced density matrix in a smaller projected Hilbert space is fully characterized. 
For the problem at hand, we trace out all modes except the
qubit degree of freedom together with an {\it arbitrary} bath 
mode with given quantum number $k$; this defines a spin and $k$-mode excluded environment 
denoted ``$\mathrm{env/spin}+k$''. The reduced ground state density matrix in the joint qubit and $k$-mode subspace reads
\begin{equation}
\label{rhospink}
\rho_{\mathrm{spin}+k} = \mathrm{Tr}_{\mathrm{env/spin}+k} |GS\big>\big<GS|.
\end{equation}
We focus here on the off-diagonal part (with respect to the qubit axis
of quantization) of the Wigner distribution 
associated to this density matrix as a function of the 
classical displacement $X$. We expect on physical grounds that this component will be most sensitive 
to the antipolaronic correlations. Its standard definition is~\cite{Raimond}
\begin{equation}
W_{\sigma^+}^{(k)}(X) \!=\!\! \int \!\!\! \frac{\mathrm{d^2}\lambda}{\pi^2}\;
e^{X(\lambda-\bar{\lambda})} 
\mathrm{Tr}_{\mathrm{spin}+k} \left[ e^{\lambda \akd-\bar{\lambda}\ak}
\sigma^+ \rho_{\mathrm{spin}+k} 
\right] ;
\label{WignerSigmaPlus}
\end{equation}
see Methods for the NRG implementation and Supplementary Information 
for discussion of the spin-diagonal part of the Wigner distribution,
which emphasizes instead the polaronic part of the total wavefunction.
%
From the two-polaron trial state (\ref{trial}) and
equation~(\ref{WignerSigmaPlus}), it is straightforward to
find the form of the Wigner function in the regime of strong
dissipation ($\alpha>0.5$):
\begin{eqnarray}
\label{decomposition}
W_{\sigma^+}^{(k)}(X) &\approx& \frac{p}{\pi}
e^{-\frac{1}{2}\sum_{q\neq k} (f^{\mathrm{pol.}}_{q}+f^{\mathrm{anti.}}_{q})^2} \\
\nonumber
&\times&
\Big[
e^{-2\big(X-\frac{f^{\mathrm{pol.}}_{k}-f^{\mathrm{anti.}}_{k}}{2}\big)^2}
+ e^{-2\big(X+\frac{f^{\mathrm{pol.}}_{k}-f^{\mathrm{anti.}}_{k}}{2}\big)^2}\Big].
\end{eqnarray}

\begin{figure}[t]
\includegraphics[width=0.9\linewidth]{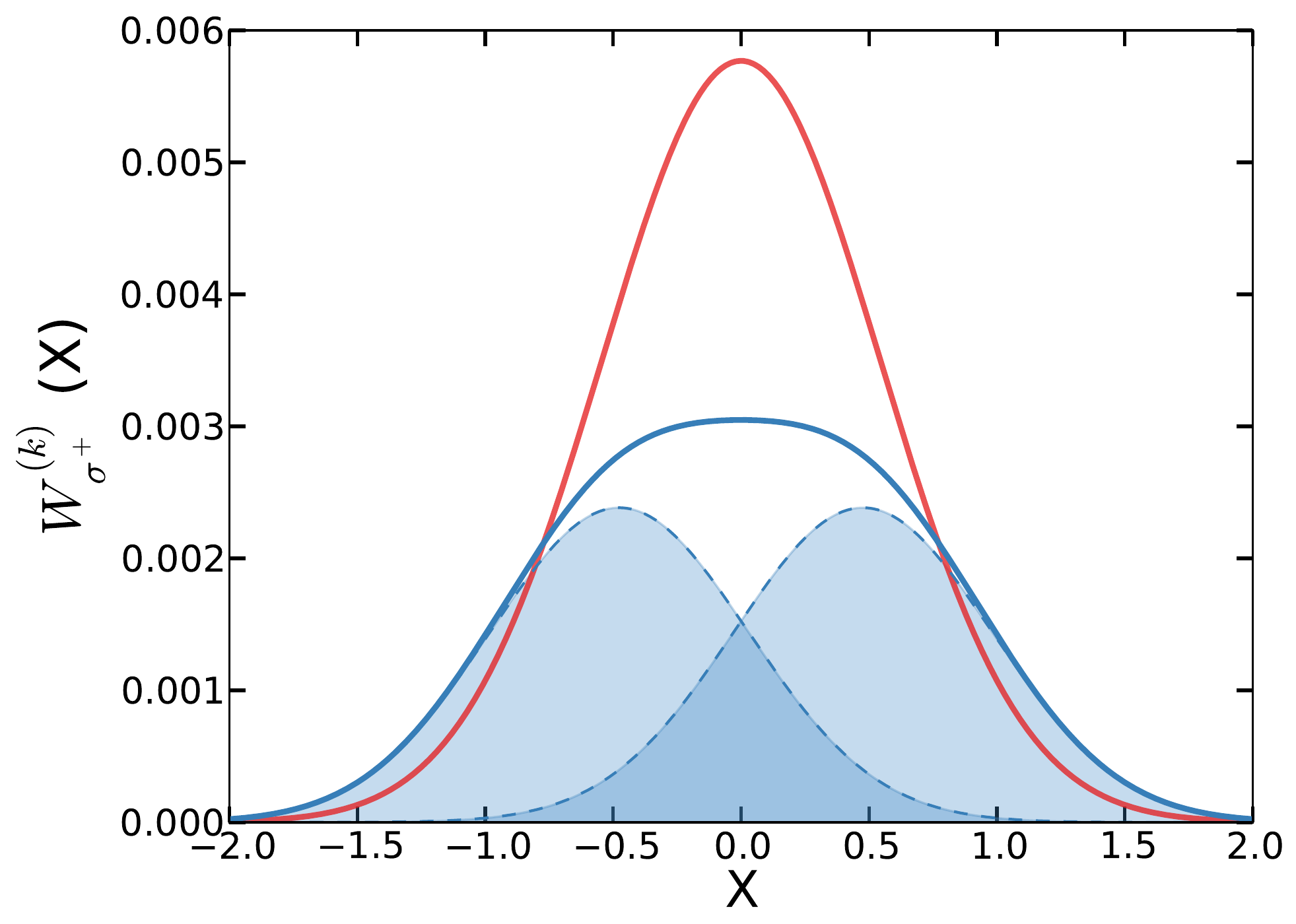}
\caption{{\bf Quantum tomography in the many-mode case.}
Transverse Wigner distribution defined in Eq.~(\ref{WignerSigmaPlus}), as obtained from 
NRG, for two different modes, one at high energy $\w_k\gg\Delta_R$ (top red curve) and 
the other at intermediate energy $\w_k\gtrsim \Delta_R$ (bottom blue curve). A decomposition of the intermediate 
energy case into two shifted Gaussians is performed (thin blue lines) according to Eq.~(\ref{decomposition}),
showing the emergence of antipolaron correlations. [Parameters  
$\alpha=0.8$ and $\Delta/\w_c=0.01$.] 
}
\label{ManyModesWigner}
\end{figure}

For high energy modes $\w_k \sim \w_c$,
$W_{\sigma^+}^{(k)}(X)$ should show a single peak centered around $X=0$,
as both polarons adiabatically follow the spin tunneling so that $f^{\mathrm{anti.}}_{k}$ becomes 
close to the polaron displacement $f^{\mathrm{pol.}}_{k}$. 
For modes of lower energy, antipolaron displacements emerge, and 
the peak separates into two lobes with displacements 
$\pm[f^{\mathrm{pol.}}_{k}-f^{\mathrm{anti.}}]\simeq \pm2f^{\mathrm{pol.}}_{k}$. 
These simple predictions of the two-polaron variational
state are clearly seen in the numerical NRG result in 
Fig.~\ref{ManyModesWigner}, strongly supporting the existence of
non-adiabatic oscillator states in the environment.

\begin{figure}[t]
\includegraphics[width=0.9\linewidth]{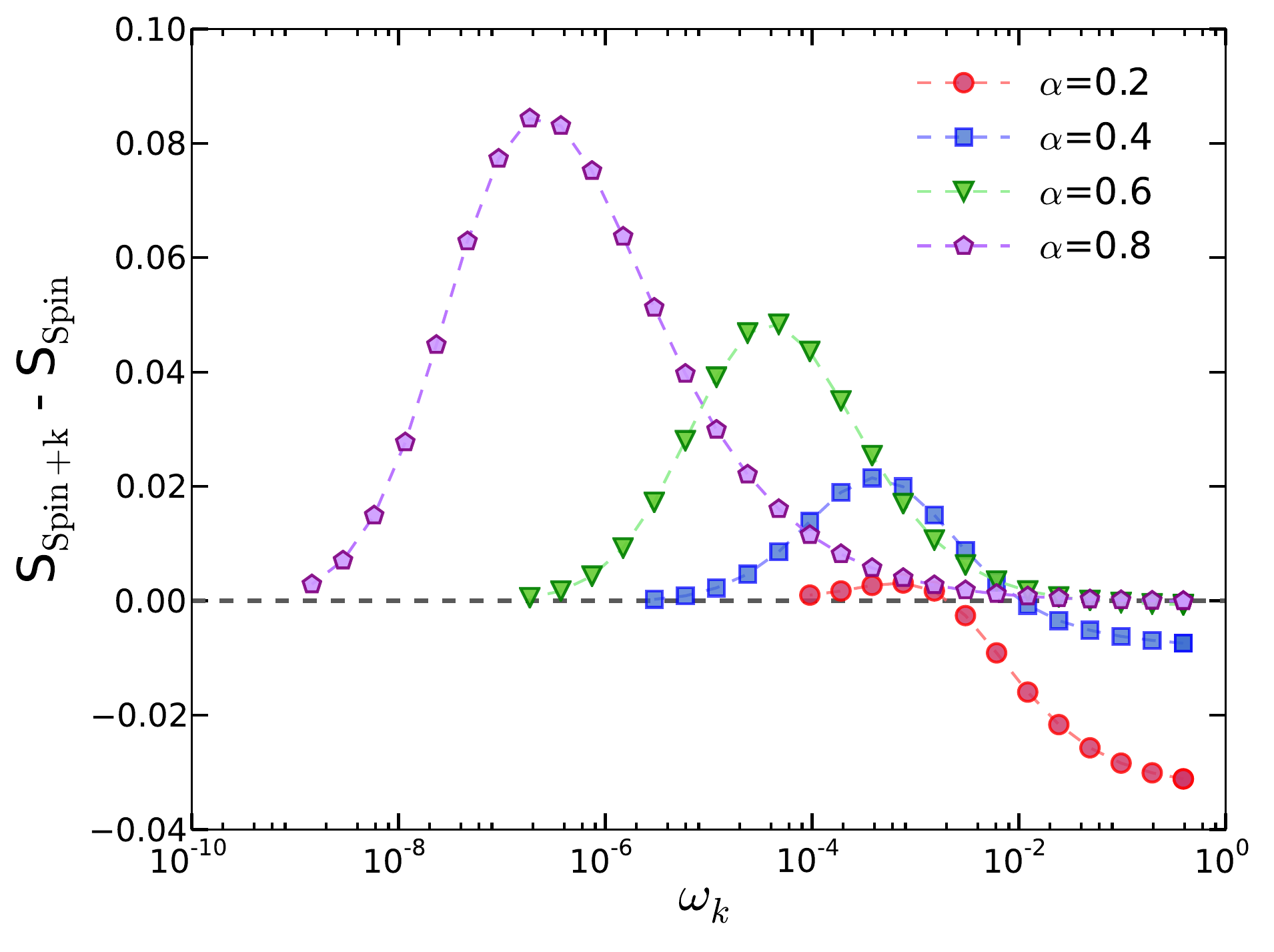}
\caption{{\bf Joint entanglement entropy in the many-mode case.}
The joint entanglement entropy of the qubit and a given $\w_k$ mode,
defined by Eq.~(\ref{Sspink}),
is calculated with NRG for $\Delta/\w_c=0.01$. 
The entropy of the qubit alone is subtracted. Negative correlations for
$\alpha<0.5$ are related to the Silbey-Harris one-polaron state, 
while the strong positive peak can only be accounted for by entanglement among
the modes of the bath. This excess entropy is, then, a sensitive measure of the
subtle non-classical correlations among the bath modes that are generated by the
coupling to the qubit.}
\label{ManyModesSAB}
\end{figure}

We finally wish to assess more directly the entanglement among the
environmental states that is suggested by the antipolaron ansatz Eq.\,(4).
For this purpose, we consider the entanglement entropy $S_{\mathrm{spin+k}}$
of the joint spin and $k$-mode subsystem with respect to the 
other modes of the bath:
\begin{equation}
\label{Sspink}
S_{\mathrm{spin+k}}= - \mathrm{Tr}_{\mathrm{spin}+k}
\left[ \rho_{\mathrm{spin}+k} \log \rho_{\mathrm{spin}+k} \right],
\end{equation}
with the reduced density matrix defined in Eq.~(\ref{rhospink}). We also
introduce the spin entanglement entropy 
$S_{\mathrm{spin}}= -\mathrm{Tr}_{\mathrm{spin}} \left[ \rho_{\mathrm{spin}} 
\log \rho_{\mathrm{spin}} \right]$ with $\rho_{\mathrm{spin}} = 
\mathrm{Tr}_{k} [\rho_{\mathrm{spin}+k}]$.
The difference between these two quantities can be computed from the NRG (see
Methods) and is plotted as a function of mode frequency in 
Fig.~\ref{ManyModesSAB}. 
For small dissipation, $\alpha<0.5$, this entropy difference is mostly negative,
as expected from the correlations built into the Silbey-Harris state, which
consist only of non-entangled environmental states within each spin-projected 
component of the wavefunction (see Supplementary Information). 
In contrast, at strong dissipation, this entropy difference becomes 
positive and shows a strikingly large enhancement near the scale $\Delta_R$.
The excess entanglement entropy, above that of the spin alone, comes from
entanglement within the bath of oscillators. This is a sensitive signature,
then, that the spin projected wavefunction is not simply a product of oscillator
states as in the SH ansatz but rather involves substantial entanglement. 
The nature of this entanglement in the simpler two-mode case is explored further 
in the Supplementary Information. 
Note especially the large energy window
where the entropy peak develops: the excess entanglement spreads from
low to high frequency modes due to the massive entangling power of the
spin tunneling operator $K_+$ discussed above in the polaron basis
$\widetilde{H}$ of Eq.~(\ref{ham}). The existence of inter-mode bosonic correlations on a
wide energy range makes also an interesting connection to the underlying 
(although  hidden in the spin-boson model) 
fermionic Kondo physics~\cite{Leggett,LeHur}.


In conclusion, we have shown how antipolaron contributions emerge in the
ground state wavefunction of the spin-boson model, causing non-classical 
Schr\"odinger cat-like environmental states. The approach here can be used 
as a general framework to expand and rationalize many-body wavefunctions in 
strongly interacting open quantum systems. 
Experimentally, proposals to realize the strongly dissipative spin
boson model using a superconducting qubit coupled to arrays of Josephson
junctions have been made very recently~\cite{LeHur,BallesterSolano,Goldstein}.
The recent progress in quantum tomography of superconducting
qubits~\cite{Hofheinz,Eichler} raises thus the challenge to measure in such
setups the massive entanglement of environment oscillators
that was unveiled here. The present work offers several directions for future
research, especially the generalization to time-dependent phenomena such as the
study of quantum quenches and spin dynamics at strong dissipation, where standard
(weak-coupling) Bloch-Redfield theory~\cite{BreuerPetruccione} is known to fail.

%

\begin{center}
{\bf Methods}
\end{center}

The numerical solution of the few-mode spin-boson Hamiltonian~(\ref{ham})
relies on standard diagonalisation procedures. In the case of a continuous
bath of oscillators, a different strategy is used. First, logarithmic shell
blocking of the bosonic modes onto energy intervals 
$[\Lambda^{-n-1}\w_c,\Lambda^{-n}\w_c]$ with $\Lambda=2$ is performed:
\begin{equation}
a^\dagger_n = \int_{\Lambda^{-n-1}\w_c}^{\Lambda^{-n}\w_c} \!\!\!\!\!\!\! 
\mathrm{d}k \; \akd \;.
\end{equation}
The resulting discrete Hamiltonian, which spans from arbitrarily small energy up 
to the high energy cutoff $\omega_c$, is then iteratively diagonalised according to the
Numerical Renormalization Group (NRG) algorithm~\cite{NRG-RMP08,Bulla}. The novel
part of the simulations performed for this work lies in the computation of
the Wigner distribution reduced to the joint spin and single $k$-mode subspace.
In order to implement Eqs.~(\ref{rhospink}-\ref{WignerSigmaPlus}), we first 
define arbitrary moments of the chosen oscillator $k$
of frequency $\w_k=\w_c\Lambda^{-n}$:
\begin{equation}
A_{\sigma_i;m,m'}^{(k)}=\big<GS| \sigma_i [a^\dagger_n]^m 
[a^{\phantom{\dagger}}_n]^{m'}|GS\big> 
\end{equation}
with $i=0,x,y,z$ labelling the Pauli matrices related to the spin projection (we
take $\sigma_0\equiv1$) and $m,m'$ positive integers.
Such ground state observables are readily computed within the NRG algorithm (for typically 
$0\leq m,m'<10$). 
One can then expand Eq.~(\ref{WignerSigmaPlus}) in a power series 
in $\lambda$ and $\bar\lambda$, yielding
\begin{equation}
W_{\sigma_i}^{(k)}(X) = \frac{2}{\pi} \sum_{m,m'=0}^{+\infty}
\!\!\! A_{\sigma_i;m,m'}^{(k)} \frac{(-1)^{m+m'}}{m!m'!} 
\frac{\partial^{m+m'}}{\partial X^{m+m'}} e^{-2X^2} .
\end{equation}
The Wigner distribution is now solely expressed in terms of the 
NRG-computable moments $A_{\sigma_i;m,m'}^{(k)}$.

A similar strategy is used for the computation of the entanglement
entropy~(\ref{Sspink}) from the reduced density matrix
$\rho_{\mathrm{spin}+k}$, which acts within the subspace spanned
by the qubit and a single bosonic mode $k$. We start by defining the 
joint spin and Fock projection operator 
$O_{\sigma_i;m,m'}^{(k)}= \sigma_i |m_k\big>\big<m_k'|$, so that 
matrix elements of the ground state density matrix simply read
\begin{equation}
\rho_{\sigma_i,m,m'}^{(k)} = \big<GS| O_{\sigma_i;m,m'}^{(k)} |GS\big> \;.
\end{equation}
This quantity is a ground state average, hence readily computable by
letting the operator $O_{\sigma_i;m,m'}^{(k)}$ evolve along the complete
NRG flow. The eigenvalues of the matrix $\rho_{\sigma_i,m,m'}^{(k)}$ 
allow one, finally, to obtain the desired entanglement entropy.

A last new 
piece of our work is the multipolaron generalization 
of the previous single-polaron trial state~\cite{Silbey,HarrisSilbey85}; this is
key for capturing easily the emergent non-adiabatic physics at strong dissipation. 
The variational method is straightforwardly implemented in the few-mode case by
minimizing the average Hamiltonian~(\ref{ham}) while using the double-polaron
ansatz~(\ref{trial}). In the many mode case, despite having two sets of unknown
functions, $f^{\mathrm{pol.}}_{k}$ and $f^{\mathrm{anti.}}_{k}$, labeled by the
continuous momentum $k$, one can show that their form as a function of $k$ is
uniquely fixed from the variational principle, leaving a finite set of effective
parameters to be determined (see Supplementary Information). 
One finds that the displacement associated with the first polaron follows qualitatively the standard behavior 
$f^{\mathrm{pol.}}_{k}=0.5g_k/(\w_k+\Delta_R)$ known from SH theory
\cite{Silbey,HarrisSilbey85}, with some quantitative deviations due to the feedback of the  
antipolaronic state $f^{\mathrm{anti.}}_{k}$. The latter takes the approximate
form 
$f^{\mathrm{anti.}}_{k}\simeq f^{\mathrm{pol.}}_{k} .
\frac{\w_k-\Omega}{\w_k+\Omega}$ with a new energy scale $\Omega$
that controls the crossover from non-adiabatic to adiabatic
behavior as a function of mode energy (see Supplementary Information for the
complete expression). The antipolaronic (non-adiabatic) character at low energy 
of the second contribution in the trial state~(\ref{trial}) is thus automatically 
guaranteed by the variational principle.

\begin{center}
{\bf Acknowledgements}
\end{center}

S.B., S.F., and H.U.B. thank the Fondation Nano\-sciences de Grenoble for
funding under RTRA contract CORTRANO. A.N. thanks Imperial College for support.
A.W.C. acknowledges support from the Winton Programme for the Physics of
Sustainability.
The work at Duke was supported by US DOE, Division of Materials Sciences and
Engineering, under Grant No. DE-SC0005237.

\setcounter{figure}{0}
\setcounter{table}{0}
\setcounter{equation}{0}

\onecolumngrid

\vspace{1.0cm}
\begin{center}
{\bf \large Supplementary information for \\ ``Unveiling environmental entanglement
in strongly dissipative qubits"}
\vspace{0.5cm}\\

\begin{quote}
{
\small

We present here additional technical details and extra results, supporting
the multi-polaronic description of the many-body ground state of the 
spin-boson model at strong coupling. We first consider the general two-polaron
variational formalism for an arbitrary number of modes. The ground state energy
and wavefunctions are then investigated for a wide range of parameter in the
single-mode (Rabi) model, highlighting the emergence of antipolaron correlations
and the possible breakdown of the single-polaron Silbey-Harris ansatz. The 
two-mode Rabi model is afterwards considered, with the emphasis on entropic
issues, which provide interesting signatures of environmental entanglement. 
The need for additional antipolaronic contributions in the
wavefunction is also discussed. Finally, the continuous spin-boson model is further 
explored, with detailed derivations of the Wigner functions pertaining to
the reduced qubit and single mode Hilbert space, as well as extra comparisons between
Numerical Renormalization Group simulations and the variational technique.
}
\end{quote}

\end{center}

\section{General two-polaron variational formalism}

\subsection{Energetics}

We consider the unbiased spin-boson model~\cite{LeggettSI,WeissSI}, as defined by the Hamiltonian (1) of
the main text: \begin{equation}\label{Hspinboson}
H=\frac{\Delta}{2}\sigma_x+\sum_k\omega_k
a_k^{\dagger}a_k-\sigma_z\sum_k\frac{g_k}{2}(a_k^{\dagger}+a_k), \end{equation}
with tunneling energy $\Delta$, a set of oscillator frequencies $\omega_k$, and
system-oscillator coupling strengths $g_k$ (assumed real). Here,
$\sigma_z=\ket{\uparrow}\bra{\uparrow}-\ket{\downarrow}\bra{\downarrow}$, with
spin basis states $\ket{\downarrow}$ and $\ket{\uparrow}$, and $a_k^{\dagger}$
($a_k$) is the oscillator creation (annihilation) operator for mode $k$.
Hamiltonian~(\ref{Hspinboson}) spans the cases of few discrete modes up to a continuum
of bosonic fields, in which case the discrete $k$-sum ought to be replaced
by an integral over energy.

Our two-polaron variational ground state ansatz takes the form
\begin{equation}\label{variationalgs}
\ket{GS^{\rm 2pol.}}=\ket{\uparrow}\left[p_1\ket{+f_k^{\rm pol.}}
+p_2\ket{+f_k^{\rm anti.}}\right]-\ket{\downarrow}\left[p_1\ket{-f_k^{\rm pol.}}
+p_2\ket{-f_k^{\rm anti.}}\right],
\end{equation}
where the bosonic part of the wavefunction involves coherent states of the form
\begin{equation}
\ket{\pm f_k}=e^{\pm\sum_k f_k(a_k^{\dagger}-a_k)}\ket{0},
\end{equation}
defined as products of displaced states, where $\ket{0}$ represents all oscillators 
being in the vacuum state. The presence of a $\mathbb{Z}_2$ symmetry, namely 
($|\uparrow\big>\to|\downarrow\big>$, $|\downarrow\big>\to|\uparrow\big>$, $a_k\to-a_k$), 
and the need for minimizing the spin tunneling energy enforces the chosen relative sign
between the up and down components of the ground state wavefunction in
Eq.~(\ref{variationalgs}).
Both functions $f_k^{\rm pol.}$ and $f_k^{\rm anti.}$ are
taken as free parameters, and will be varied to minimise the total ground state energy
$E=\bra{GS^{\rm 2pol.}}H\ket{GS^{\rm 2pol.}}$. In contrast to the usual
Silbey-Harris state (for which $p_2=0$)~\cite{SilbeySI,Chin_SubohmSBM11SI,NazirSI}, this more flexible 
ansatz allows for the possibility of a superposition of variationally determined displaced 
oscillator states associated with each spin projection.

Normalisation of $\ket{GS^{\rm 2pol.}}$ implies the condition
\begin{equation}\label{normcondition}
2p_1^2+2p_2^2+4p_1p_2e^{-\frac{1}{2}\sum_k(f_k^{\rm pol.}-f_k^{\rm anti.})^2}=1,
\end{equation}
while the variational ground state energy is given by
\begin{eqnarray}\label{evargs}
E=\bra{GS^{\rm 2pol.}}H\ket{GS^{\rm 2pol.}}&=&-\Delta\left(p_1^2e^{-2\sum_k(f_k^{\rm pol.})^2}
+p_2^2e^{-2\sum_k(f_k^{\rm anti.})^2}
+2p_1p_2e^{-\frac{1}{2}\sum_k(f_k^{\rm pol.}+f_k^{\rm anti.})^2}\right)\nonumber\\
&&\:+2\sum_k\omega_k\left(p_1^2(f_k^{\rm pol.})^2+p_2^2(f_k^{\rm anti.})^2
+2p_1p_2f_k^{\rm pol.}f_k^{\rm anti.}e^{-\frac{1}{2}\sum_k(f_k^{\rm pol.}-f_k^{\rm anti.})^2}\right)\nonumber\\
&&\:{-}2\sum_kg_k\left(p_1^2f_k^{\rm pol.}+p_2^2f_k^{\rm anti.}
+p_1p_2(f_k^{\rm pol.}+f_k^{\rm anti.})e^{-\frac{1}{2}\sum_k(f_k^{\rm pol.}-f_k^{\rm anti.})^2}\right).
\end{eqnarray}
In the limit that $p_2\rightarrow0$ (and so $p_1\rightarrow1/\sqrt{2}$) we recover the 
Silbey-Harris variational ground state energy, 
\begin{equation}\label{eSHgs}
E_{\rm{SH}}=-\frac{\Delta}{2}e^{-2\sum_k(f_k^{\rm pol.})^2}+\sum_k\omega_k (f_k^{\rm pol.})^2-\sum_kg_kf_k^{\rm pol.},
\end{equation}
while further setting $\alpha_k=g_k/(2\omega_k)$ in Eq.~(\ref{eSHgs}) gives the
(non-variationally optimal) bare polaron ground state energy
\begin{equation}\label{epolgs}
E_{\rm{POL}}=-\frac{\Delta}{2}e^{-\frac{1}{2}\sum_kg_k^2/\omega_k^2}-\sum_k\frac{g_k^2}{4\omega_k}.
\end{equation}
Going back to the two-polaron variational state of Eq.~(\ref{variationalgs}), we find that the ground 
state coherence is given by
\begin{equation}\label{gscoh}
\langle\sigma_x\rangle=-2\left(p_1^2e^{-2\sum_k(f_k^{\rm pol.})^2}
+p_2^2e^{-2\sum_k(f_k^{\rm anti.})^2}+2p_1p_2e^{-\frac{1}{2}\sum_k(f_k^{\rm pol.}+f_k^{\rm anti.})^2}\right),
\end{equation}
while the magnetisation $\langle\sigma_z\rangle=0$ by symmetry in absence of
magnetic field along $\sigma_z$ (unless one enters the polarized phase at $\alpha>1$ in 
the ohmic spin-boson model).

\subsection{Variational displacements}

The two sets of displacements $f_k^\mathrm{pol.}$ and $f_k^\mathrm{anti.}$ are
variationally determined from the total energy $E$ of Eq.~(\ref{evargs}) according 
to $\partial E/\partial f_k^\mathrm{pol.}=0$ and $\partial E/\partial f_k^\mathrm{anti.}=0$, 
which gives the closed form:
\begin{eqnarray}
\label{Supfkpol}
f_k^\mathrm{pol.} &=& \frac{g_k}{2}
\frac{A_1 (p_2^2 \w_k + \Delta_2)-
A_2[p_1p_2\big<f_k^\mathrm{pol.}|f_k^\mathrm{anti.}\big>+\Delta_{12}]}
{(p_1^2\w_k+\Delta_1)(p_2^2\w_k+\Delta_2)-
[p_1p_2\big<f_k^\mathrm{pol.}|f_k^\mathrm{anti.}\big>+\Delta_{12}]^2}\\
f_k^\mathrm{anti.} &=& \frac{g_k}{2}
\frac{A_2 (p_1^2 \w_k + \Delta_1)-
A_1[p_1p_2\big<f_k^\mathrm{pol.}|f_k^\mathrm{anti.}\big>+\Delta_{12}]}
{(p_1^2\w_k+\Delta_1)(p_2^2\w_k+\Delta_2)-
[p_1p_2\big<f_k^\mathrm{pol.}|f_k^\mathrm{anti.}\big>+\Delta_{12}]^2},
\end{eqnarray}
which is valid for an arbitrary number of oscillator modes.
Hence the generic $k$-dependence of the displacement is fully constrained by the
variational principle, which leaves a finite set of effective parameters to
be determined self-consistently according to:
\begin{eqnarray}
\Delta_1 &=& \Delta p_1^2 \big<-f_k^\mathrm{pol.}|f_k^\mathrm{pol.}\big>
+ \frac{\Delta}{2} p_1p_2 \big<-f_k^\mathrm{pol.}|f_k^\mathrm{anti.}\big>
+ p_1p_2 (-\tilde{\w}+\tilde{g}) \big<f_k^\mathrm{pol.}|f_k^\mathrm{anti.}\big>\\
\Delta_2 &=& \Delta p_2^2 \big<-f_k^\mathrm{anti.}|f_k^\mathrm{anti.}\big>
+ \frac{\Delta}{2} p_1p_2 \big<-f_k^\mathrm{pol.}|f_k^\mathrm{anti.}\big>
+ p_1p_2 (-\tilde{\w}+\tilde{g}) \big<f_k^\mathrm{pol.}|f_k^\mathrm{anti.}\big>\\
\Delta_{12} &=& 
\frac{\Delta}{2} p_1p_2 \big<-f_k^\mathrm{pol.}|f_k^\mathrm{anti.}\big>
+ p_1p_2 (\tilde{\w}-\tilde{g}) \big<f_k^\mathrm{pol.}|f_k^\mathrm{anti.}\big>\\
A_{1} &=& p_1^2+ 2p_1p_2 \big<f_k^\mathrm{pol.}|f_k^\mathrm{anti.}\big>\\
A_{2} &=& p_2^2+ 2p_1p_2 \big<f_k^\mathrm{pol.}|f_k^\mathrm{anti.}\big>\\
\tilde{\w} &=& \sum_k \w_k f_k^\mathrm{pol.}f_k^\mathrm{anti.}\\
\tilde{g} &=& \sum_k g_k [f_k^\mathrm{pol.}+f_k^\mathrm{anti.}].
\end{eqnarray}
The one-polaron Silbey-Harris displacement
$f_k^\mathrm{SH}=0.5g_k/\big[\w_k+\Delta\big<-f_k^\mathrm{pol.}|f_k^\mathrm{pol.}\big>\big]$ 
is trivially recovered from Eq.~(\ref{Supfkpol}) by letting $p_2=0$.
One can also check that $f_k^\mathrm{anti.} \simeq f_k^\mathrm{pol.}$ for
$k\to\infty$, while $f_k^\mathrm{anti.}\simeq -f_k^\mathrm{pol.}$ for
$k\to0$ in the limit of strong dissipation. Thus the antipolaron displacement satisfies 
the expected adiabatic/non-adiabatic crossover as a function of energy $\w_k$, and
this physical picture is naturally incorporated in the variational theory.

\section{Single-mode Rabi model: checking the wavefunction}

It is illustrative to consider the simplified case of a single-mode within the
environment, namely the Rabi model (see Ref.~\onlinecite{BraakSI} and references
therein). In this situation, the model Hamiltonian may be diagonalised
straightforwardly (numerically) and so the regimes of validity of our
polaron-antipolaron ansatz, as well as of the Silbey-Harris and non-optimal
bare polaron states, may be assessed. The Hamiltonian now becomes
\begin{equation}\label{Hspinbosonsingle}
H_1=\frac{\Delta}{2}\sigma_x+\omega_1 a_1^{\dagger}a_1-\sigma_z\frac{g}{2}(a_1^{\dagger}+a_1),
\end{equation}
with ground-state ansatz
\begin{equation}\label{variationalgssingle}
\ket{GS_1^{\rm 2pol.}}=\ket{\uparrow}\left[p_1\ket{+f_1^{\rm pol.}}
+p_2\ket{+f_1^{\rm anti.}}\right]-\ket{\downarrow}\left[p_1\ket{-f_1^{\rm pol.}}
+p_2\ket{-f_1^{\rm anti.}}\right],
\end{equation}
where $\ket{\pm f_1}=e^{\pm f_1(a_1^{\dagger}-a_1)}\ket{0}$. To optimise the
state, we minimise the variational ground state energy $E=\bra{GS_1^{\rm
2pol.}}H_1\ket{GS_1^{\rm 2pol.}}$ numerically, subject to the normalisation
constraint $2p_1^2+2p_2^2+4p_1p_2e^{-\frac{1}{2}(f_1^{\rm pol.}-f_1^{\rm
anti.})^2}=1$. 
The Silbey-Harris state is again obtained by letting $p_2\rightarrow0$, such that 
\begin{equation}\label{SHgssingle}
\ket{GS_1^{\rm SH}}=\frac{1}{\sqrt{2}}\left[\ket{\uparrow}\ket{+f_1^{\rm SH}}
-\ket{\downarrow}\ket{-f_1^{\rm SH}}\right],
\end{equation}
and there is only a single displaced oscillator associated with each spin.
Minimisation of $E^{\rm SH}=\bra{GS_1^{\rm SH}}H_1\ket{GS_1^{\rm SH}}$, leads to
a self-consistent equation for the optimised displacement, $f_1^{\rm
SH}=g/[2(\Delta e^{-2(f_1^{SH})^2}+\omega_1)]$. The non-optimal bare polaron state has the same
form as Eq.~(\ref{SHgssingle}), but with the displacement fixed at $f_1=g/(2\omega_1)$.
\begin{figure}[!ht]
\begin{center}
\includegraphics[width=0.85\textwidth]{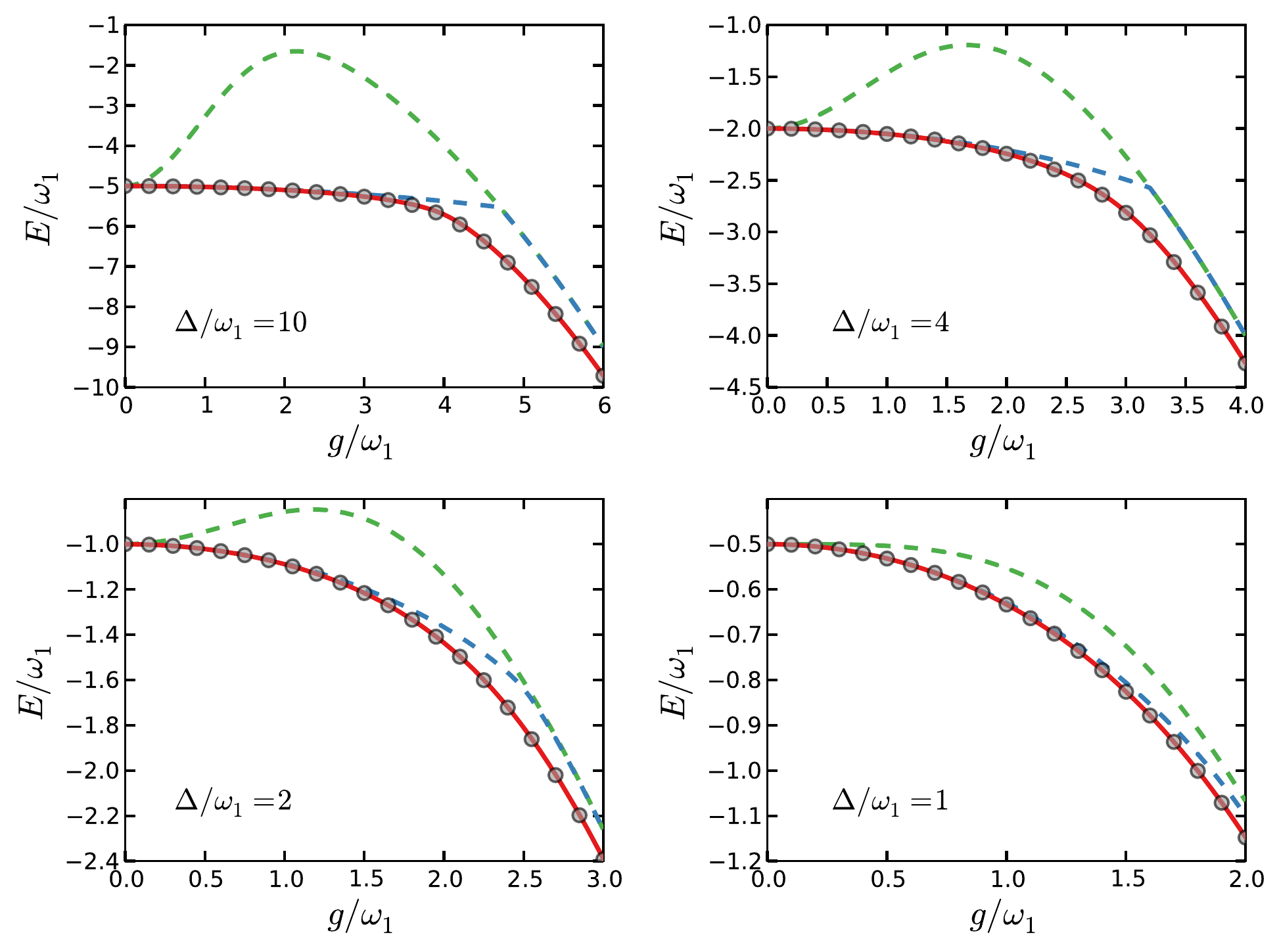}
\caption{{\bf Ground state energy of the Rabi model.} The single-mode spin boson
model is considered as a function of spin-oscillator coupling strength, and the
ground state energy calculated from our variational two-polaron ansatz (red 
solid curves), the Silbey-Harris one-polaron variational state (blue dashed
curves), and from the bare polaron ground state (green dashed curves). Exact energies 
calculated by numerically diagonalising the Hamiltonian
(for a basis of 200 states) are shown as grey dots. The values of $\Delta/\omega_1$
used are shown on each plot. For all plots, $\Delta/\omega_1\geq1$, which is the regime
in which we expect our variational state to outperform the Silbey-Harris or bare
polaron treatments.} 
\label{eground1}
\end{center}
\end{figure}

In Fig.~\ref{eground1} we plot the dimensionless ground state energy $E/\omega_1$ determined
from our polaron-antipolaron variational ansatz as a function of the
dimensionless spin-oscillator coupling strength $g/\omega_1$, and compare with the 
results from an exact numerical diagonalisation of the model, and from Silbey-Harris and polaron theories
(Eqs.~(\ref{eSHgs}) and (\ref{epolgs}) in the single mode case, respectively).
In this figure, $\Delta/\omega_1\geq1$ for all plots, and so we would expect
standard polaron theory to break down in this regime, since the full oscillator
displacement is no longer appropriate. From the dashed curves, this can indeed
be seen to be the case, and polaron theory may even predict the incorrect trend
as $g/\omega_1$ increases. Silbey-Harris theory fixes this problem to a certain
extent (at least at small $g/\omega_1$, see dashed-dotted curves), though again
runs into problems as the coupling strength increases, deviating from the
numerically-exact results, and even more worryingly predicting discontinuous
behaviour in the ground-state energy at certain values of $g/\omega_1$. Our
polaron-antipolaron variational ansatz, however, predicts ground-state energies
in almost {\it perfect agreement} with the numerical results for all possible coupling
strengths. Furthermore, the discontinuous behaviour seen in Silbey-Harris
theory is removed in this more flexible variational state.

The failure of the single-polaron theories can be interpreted by analysing, in position 
space, the oscillator states associated with the spin projetions $\ket{\downarrow}$ 
and $\ket{\uparrow}$ in the model ground state ($\phi^0_{\downarrow}(x)$ and $\phi^0_{\uparrow}(x)$,
respectively), where we take the ground state to have the form
$\ket{GS_1}=\ket{{\phi^0_{\uparrow}}}\ket{\uparrow}+\ket{\phi^0_{\downarrow}}\ket{\downarrow}$,
i.e. we absorb any normalisation factors and minus signs into the oscillator
states, $\ket{\phi^0_{\downarrow}}$ is thus negative using this convention.
\begin{figure}[!ht]
\begin{center}
\includegraphics[width=0.85\textwidth]{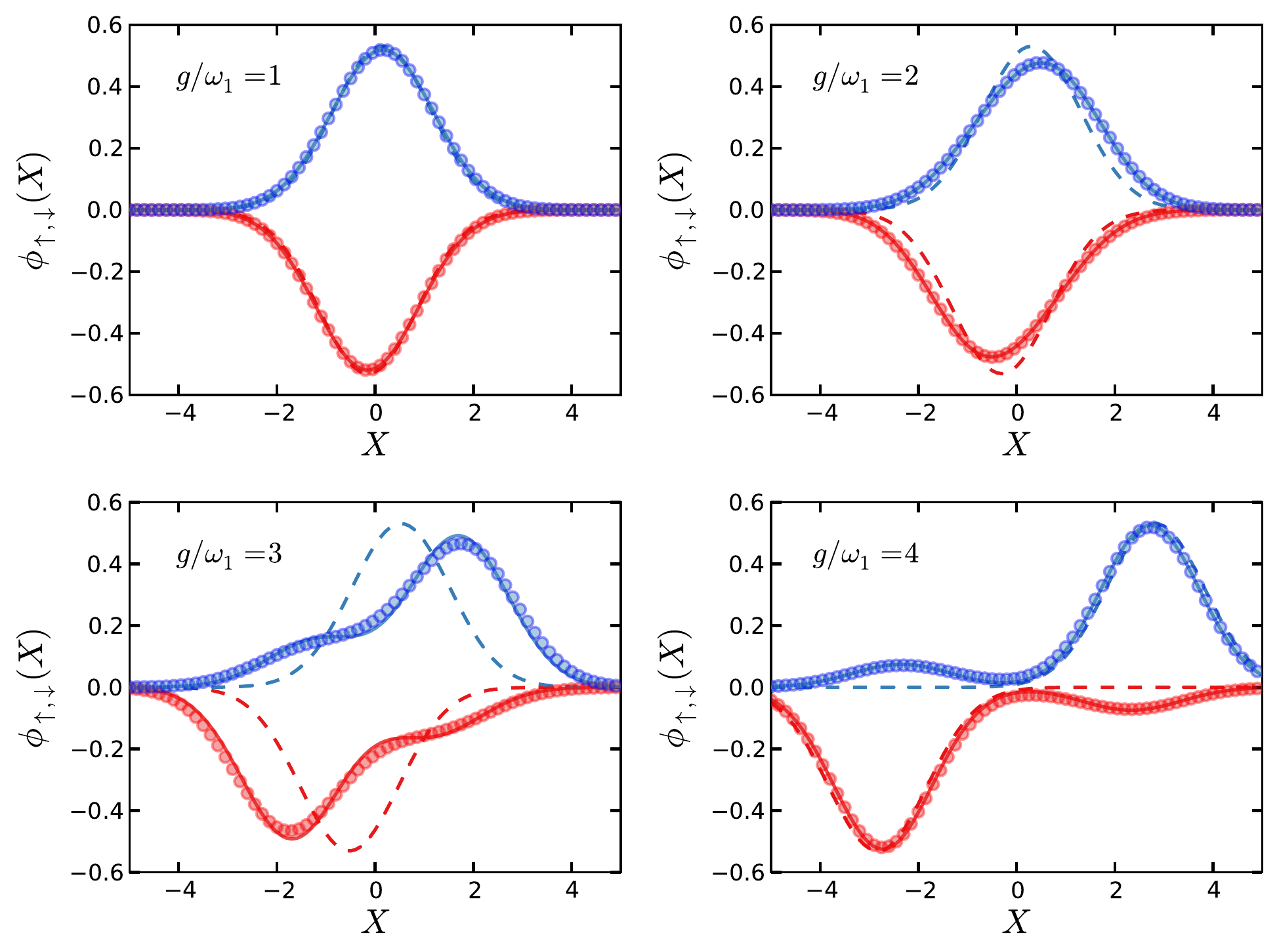}
\caption{{\bf Ground state wavefunctions of the Rabi model.} Oscillator states $\phi^0_{\downarrow}(x)$ (red) and
$\phi^0_{\uparrow}(x)$ (blue) plotted as a function of position for
$\Delta/\omega_1=4$. Shown are the predictions of our variational ground state
(solid curves), of the Silbey-Harris ground state (dashed-dotted curves), and
exact calculations from a numerical diagonalisation of the full Hamiltonian
(points).} 
\label{osc2}
\end{center}
\end{figure}
Shown in Fig.~\ref{osc2} are thus $\phi^0_{\downarrow}(x)$ (red) and
$\phi^0_{\uparrow}(x)$ (blue) as a function of position, calculated from our
polaron-antipolaron ground state (solid curves), from the Silbey-Harris ground
state (dashed-dotted curves), and from a numerical diagonalisation of the full
Hamiltonian (points). We take $\Delta/\omega_1=4$ here as a representative
example. For $g/\omega_1\sim1$, the displacement from $x=0$ is found to be
fairly small, $|f|<g/(2\omega_1)$, which is why the full polaron approach
fails, see Fig.~\ref{eground1}). The correct displacements can be captured by the 
Silbey-Harris theory and as well by the more flexible two-polaron ansatz presented in
Eq.~(\ref{variationalgssingle}). However, as the coupling strength increases
further, we can see that the double displacement nature of the oscillator states starts to
become extremely important. For example, at $g/\omega_1=3$ we observe that the
Silbey-Harris state is completely unable to reproduce the correct oscillator
wavefunctions, due to the restriction to a single displacement associated with each 
spin state. In fact, for
these parameters, the displacements obtained by the Silbey-Harris approach are
much too small, and reproduce none of displacements seen in our
polaron-antipolaron ansatz, which itself matches the numerical solution very
well. Finally, as the coupling strength is increased further, the Silbey-Harris
displacements eventually ``jump" to those of the full polaron transformation,
which then captures the dominant displacements in the exact states quite well
(see the $g/\omega_1=4$ plot ), but of course completely misses the smaller
displacements in the opposite direction, which are still captured extremely well
by our ansatz. Hence, the obtained ground state energy is still lower in our
ansatz (and in the numerical diagonalisation) than from the Silbey-Harris state.
The theories will eventually converge at larger coupling, however, when
associating a single (polaron) displacement with each spin state finally becomes
a good description. 

\section{Two-mode Rabi model: entanglement entropy}

\begin{figure}[!ht]
\begin{center}
\includegraphics[width=18cm]{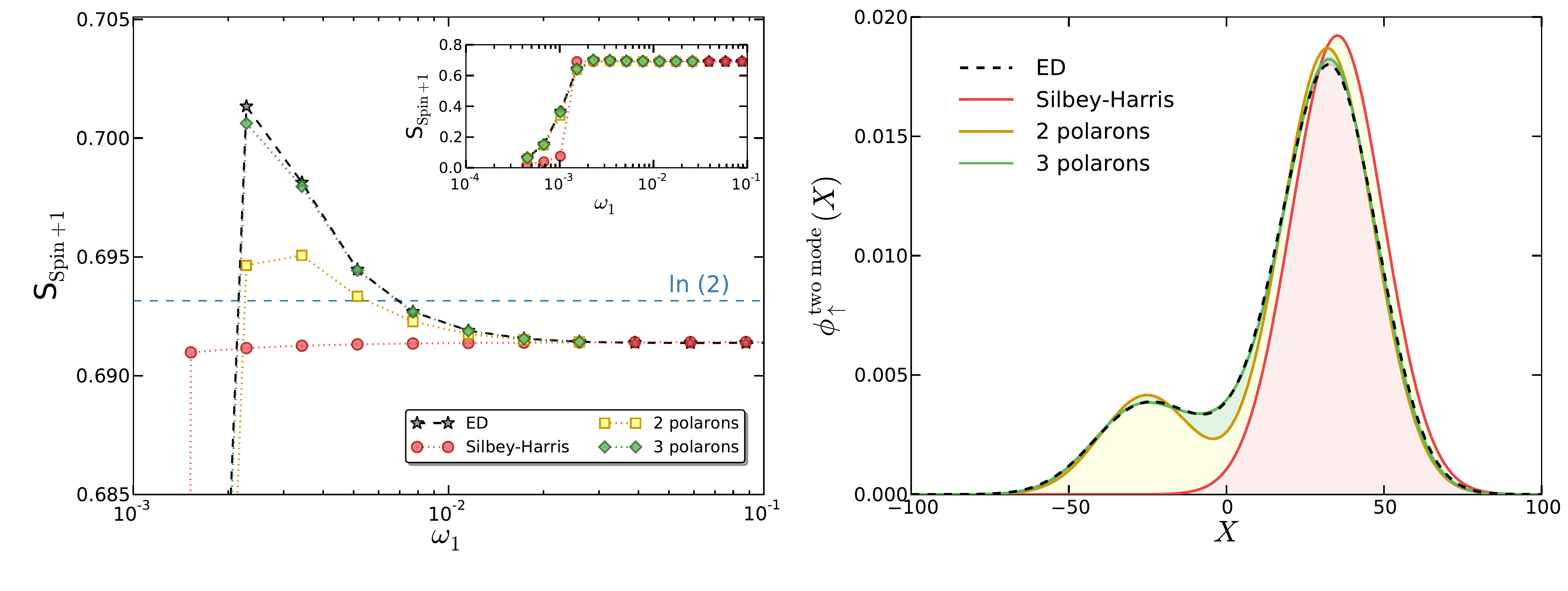}
\caption{{\bf Joint spin-mode entropies and oscillator wavefunctions for the
two-mode spin-boson model}. {\bf A}.  Joint spin-mode entropy $S_{\mathrm{Spin}+1}$ for a
two-mode environment. For all these data $\omega_{2}=1.05\omega_{1}$,
$g_{1}=g_{2}=2.5\omega_{1}$ and $\Delta=0.01$, sweeping the frequencies
$\omega_1$ of the first mode.
Plot shows results for (top to bottom) exact diagonalisation (grey stars), three-polaron ansatz 
(green diamonds), two-polaron ansatz (yellow squares) and
one-polaron Silbey-Harris theory (red dots). The horizontal line
indicates the maximum entropy of a fully-mixed spin state ($\ln(2)$). The main
panel is a close-up on the entropy peak occuring near the resonance frequency,
as discussed in the text, while the inset shows the whole entropy and frequency
range.
{\bf B}. 
Spin up-projected two-mode oscillator wave functions along the diagonal coordinate
$x_{1}=x_{2}$ for the same parameters as the left panel, computed for 
$\omega_{1}=0.0023$ (namely at the peak position of the exact diagonalisation
entropy curve in {\bf A}). Results are shown for ground states obtained by exact
diagonalisation (dashed black line), one-polaron Silbey-Harris ansatz (solid red
line), two-polaron ansatz (solid yellow line) and three-polaron ansatz (solid green 
line). } 
\label{twomodeent}
\end{center}
\end{figure}

In order to understand the non-monotoic behaviour of the joint spin-mode entropy
$S_{\mathrm{Spin}+k}$ results presented for the continuum environment in Fig.~5 of the
main text, we consider in this appendix the entanglement \emph{between} modes in the
environment for the simpler case of a two-mode environment. The Hamiltonian for
this case, the same used to generate the two-mode results in the main text, is
given by 
\begin{equation}
H=\frac{\Delta}{2}\sigma_{x}
-\frac{\sigma_{z}}{2}\sum_{i=1,2}g_{i}(a_{i}+a^{\dagger}_{i})
+\sum_{i=1,2}\omega_{i}a^{\dagger}_{i}a_{i}.
\end{equation}   
For this two-mode environment the ground state can be found by exact
diagonalisation (ED) techniques, allowing a comparison to be made with the
predictions of our anti-polaron ansatz. Figure \ref{twomodeent}{\bf A} shows results
for $S_{\mathrm{Spin}+1}$ (we trace over mode $2$) as a function of $\omega_{1}$ for
ground state obtained by ED and three variational ans\"{a}tze: Silbey-Harris
(one-polaron), two-polaron, and three-polaron (to be discussed below) trial states. 
In all cases, a fixed relative detuning of the modes and ratio of coupling to frequency was kept, with
$\omega_{2}=1.05\omega_{1}$, $g_{1}=g_{2}=2.5\omega_{1}$. For large frequencies
$\omega_{1}\gg \Delta_{R}$, where we expect adiabatic polaron theory to describe
the state, the product state (unentangled) form of the wave functions of
oscillators $1$ and $2$ in the spin-projected states is expected to lead
$S_{\mathrm{Spin}+1}$ to be controlled only by the polaronic correlations between mode
$2$ and the spin. This is determined by the renormalisation scale $\Delta_{R}$,
which is a function of only the ratio $g_{2}/\omega_{2}$. As this is held fixed
in these simulations, we expect $S_{\mathrm{Spin}+1}$ to approach a constant value at
high frequency. This behaviour is indeed observed on all curves in Fig.
\ref{twomodeent}{\bf A}, with the relative strong coupling ($g_{2}/\omega_{2}=2.5$)
leading to an almost fully-mixed spin state.  According to Silbey-Harris theory,
which has the same wave function structure as the adiabatic polaron theory but
with different displacements, oscillators with frequencies well below
$\Delta_{R}$ should have suppressed displacements. Consequently, the
renormalisation of the spin tunneling by slow modes should be continuously
suppressed, leading to reduction of $S_{\mathrm{Spin}+1}$ as $\omega_{1}\rightarrow0$.
This behaviour is precisely what is observed for the Silbey-Harris results in
Fig. \ref{twomodeent}{\bf A}, with $S_{\mathrm{Spin}+1}$ decreasing monotonically with
decreasing frequency (though with a sharp suppression below the spin-tunneling
frequency scale).       

However, as in the multimode case, we see that the numerically exact ED ground
state shows an entropy peak (in excess of the entropy of a fully-mixed spin
state) at frequencies close to the scale $\Delta_{R}$. As was shown in the
two-mode wave function plots in the main text, the spin-projected oscillator
wavefunctions in the antipolaron regime are entangled by their non-adiabatic
responses, and these inter-mode correlations lead to the entropy peak seen in
Fig. \ref{twomodeent}{\bf A} (due to \emph{both} spin and mode $1$ states becoming
mixed when mode $2$ is traced over). As in the single mode case, this occurs in
the regime where antipolarons form, $g_{i}\approx\omega_{i}\approx\Delta_{R}$.
The absence of the antipolaron component in the Silbey-Harris theory means that
this feature cannot be described by this state. 

However, a variationally-optimised two-polaron ground state ansatz as in
Eq.~(\ref{variationalgssingle}) 
captures this peak structure in the frequency dependence of $S_{\mathrm{Spin}+1}$. To
understand qualitatively how antipolaron components in the two-polaron ansatz
generate inter-mode entanglement and, thus, the increased spin-mode entropy,
consider the (spin-up) spin-projected part of the ground state. This gives a
contribution to the total density matrix of 
\begin{equation}
\rho_{\uparrow\uparrow,1,2}=|\uparrow\rangle\langle\uparrow|
\left(p_1^2 |f_{k}^{{\rm pol.}}\rangle\langle f_{k}^{{\rm pol.}}|
+p_2^2|f_{k}^{{\rm anti.}}\rangle\langle f_{k}^{{\rm anti.}}|
+p_1p_2|f_{k}^{{\rm pol.}}\rangle\langle f_{k}^{\rm{anti.}}|
+p_1p_2|f_{k}^{{\rm anti.}}\rangle\langle f_{k}^{{\rm pol.}}|\right),
\end{equation}
where $|f_{k}^{x}\rangle=e^{\sum_{k=1,2}f_{k}^{x}(a^{\dagger}_{k}-a_{k})}|0\rangle$.
Tracing over mode $2$, the reduced state $\rho_{\uparrow\uparrow,1}$ is
 \begin{equation}
\rho_{\uparrow\uparrow,1}=|\uparrow\rangle\langle\uparrow|
\left( p_1^2 |f_{1}^{{\rm pol.}}\rangle\langle f_{1}^{{\rm pol.}}|
+p_2^2|f_{1}^{{\rm anti.}}\rangle\langle f_{1}^{{\rm anti.}}|
+p_1p_2\Phi_{22}|f_{1}^{{\rm pol.}}\rangle\langle f_{1}^{\rm{anti.}}|
+p_1p_2\Phi_{22}|f_{1}^{{\rm anti.}}\rangle\langle f_{1}^{{\rm pol.}}|\right),
\end{equation}
where $\Phi_{22}={\rm Tr}_{2}\left [|f_{2}^{{\rm pol.}}\rangle\langle
f_{2}^{\rm{anti.}}|\right ]=e^{-(f_{2}^{{\rm pol.}}-f_{2}^{{\rm anti.}})^2/2}$.
As $f_{2}^{{\rm pol.}}$ and $f_{2}^{{\rm anti.}}$ have opposite signs when mode
$2$ is in the antipolaron regime, the overlap integral $\Phi_{22}$ suppresses the
purity of the spin-projected state of mode $1$. In the extreme case where
$\Phi_{22}\approx 0$, the mixed reduced state is   

 \begin{equation}
\rho_{\uparrow\uparrow,1}\approx|\uparrow\rangle\langle\uparrow|
\left( p_1^2 |f_{1}^{{\rm pol.}}\rangle\langle f_{1}^{{\rm pol.}}|
+p_2^2|f_{1}^{{\rm anti.}}\rangle\langle f_{1}^{{\rm anti.}}|\right).
\label{mixed}
\end{equation}

This expression provides important intuition as to frequency dependence of
inter-mode entanglement. The state given in Eq. (\ref{mixed}) will have high
entropy if the antipolaron weight $p_2/p_1$ in the wave function is significant and
the states $|f_{1}^{{\rm anti.}}\rangle$ and $|f_{1}^{{\rm pol.}}\rangle$ have
weak overlap (i.e. are close to orthogonal) . We therefore expect that the
entropy peak will appear in the strong antipolaron regime (where $p_2/p_1$
and the displacements  $f_{1}^{{\rm pol.}}\approx -f_{1}^{{\rm anti.}}$ are
sizeable).
However, this also requires that $\Phi_{22}$ is also $\approx 0$, requiring that
mode $2$ is also in the strong anti-polaron regime. Our theory thus establishes
the microscopic link between the appearance of \emph{intra-mode}
non-classicality (cat-states/antipolaron) and \emph{many-body} entanglement
between such modes. For the results of Fig. \ref{twomodeent}{\bf A}, the $5\%$ detuning
between modes leads to both modes showing antipolaron features at similar
frequencies, leading to the large entropy peak. In the multimode case, we
therefore expect to find antipolaron components in the environmental
wavefunctions everywhere inside the region of positive 
$S_{\mathrm{Spin}+k}-S_{\mathrm{Spin}}$,
allowing the frequency range and position of the non-classical (cat-like)
environmental features - which could be experimentally probed through the
environmental response function - to be inferred.  We have also checked that
when the detuning between modes is made larger, such that the modes do not both
develop antipolarons for the same parameters, the entropy peak becomes much
smaller (not shown). 

Finally, we note that while the two-polaron ansatz captures the essential
physics of the entropy peak, the preservation of $\langle \sigma_{x} \rangle$ at
strong coupling, and also provides an excellent description of the shapes of the
entangled wave functions, the agreement with the ED entropy is not perfect, with
the two-polaron ansatz slightly underestimating the peak entropy, see 
Fig.~\ref{twomodeent}{\bf A}. To investigate this remaining small discrepancy, we 
have also implemented a variational three-polaron ansatz of the form:
\begin{eqnarray}
\label{trialSI}
\big|GS^{\mathrm{3pol.}}\big> &=& \big|\uparrow\big>
\otimes \left[ 
p_1\big|\!+\!f^{\mathrm{pol.}}_{k}\big> +
p_2 \big|\!+\!f^{\mathrm{anti.}}_{k}\big> +
p_{3} \big|\!+\!f^{3}_{k}\big> \right]\\
\nonumber
&-& \big|\downarrow\big> 
\otimes \left[ 
p_1\big|\!-\!f^{\mathrm{pol.}}_{k}\big> +
p_2  \big|\!-\!f^{\mathrm{anti.}}_{k}\big> +
p_{3} \big|\!-f^{3}_{k}\big> \right].
\end{eqnarray}

Figure \ref{twomodeent}{\bf B} shows the wave functions of the spin up-projected 
states of the oscillators along the diagonal coordinate
$x_{1}=x_{2}$ for the ground states obtained by ED and the two and three-polaron
ans\"{a}tze. Comparing the ED and two-polaron wavefunctions, we see that the
two-polaron wavefunctions captures the displacements and weights of the polaron and
antipolaron very well, but slightly underestimates the amplitude of the
wavefunction around the origin. For simplicity, the three-polaron solution was
determined by fixing $f_{k}^{3}=0$ and treating all other parameters
variationally. The result, shown in Fig.~\ref{twomodeent}{\bf B} has almost
perfect overlap with the ED results and gives an improved prediction for the
entropy peak in Fig.~\ref{twomodeent}{\bf A}. This result suggests that it may be
fruitful to consider a multipolaron expansion of the state in the many-mode
cases, particularly if one is interested in reproducing sensitive measures of
the many-body state structure (such as the joint entropy or other tomographic
objects) rather than the simple spin observables (which are already
well-approximated by the two-polaron results).

\section{Multi-mode spin-boson model: Wigner distributions}

\begin{figure}[!t]
\begin{center}
\includegraphics[width=0.85\textwidth]{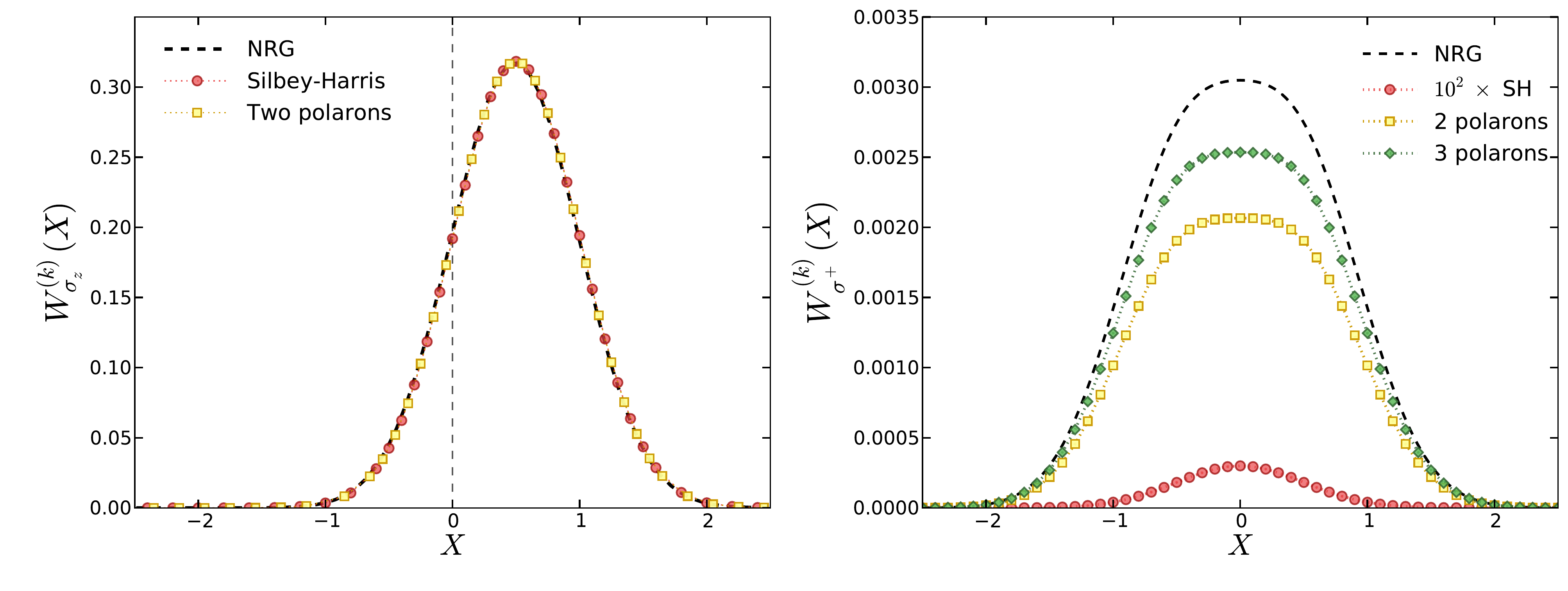}
\caption{{\bf Wigner distributions for the many-mode spin-boson model.}
{\bf A.} Diagonal Wigner distribution defined in Eq.~(\ref{SupWSz}) computed
by the NRG (dashed line) and compared to the one and two-polaron 
results found from Eq.~(\ref{SupWSzFinal}). Parameters are $\alpha=0.8$ and
$\Delta/\omega_c=0.01$. The one-polaron Silbey-Harris state is enough to account 
very well for the polaron content of the exact wavefunction.
{\bf B.} Off-diagonal Wigner distribution defined in Eq.~(\ref{SupWSplus}) computed
by the NRG (dashed line) and compared to the one and two-polaron 
results found from Eq.~(\ref{SupWSplusFinal}). 
The general magnitude and non-Gaussian
form of the Wigner distribution is only accounted for by the two-polaron state,
with a complete failure of the Silbey-Harris wavefunction (note the 100$\times$ 
magnification used to reveal its tiny contribution to the Wigner distribution). 
The remaining quantitative deviations in the two-polaron ansatz are due to the presence 
of additional antipolaronic contributions in the total wavefunction, as can be inferred 
from the computation done within a three-polaron trial state.}
\label{WignerCompare}
\end{center}
\end{figure}

We discuss here various Wigner distributions associated to the reduced
density matrix living in the subspace spanned by the qubit and one {\it given}
oscillator mode with frequency $\omega_k$. The qubit degrees of freedom
can be used for filtering out the polaron and antipolaron contributions within
the wavefunction, thanks to appropriate insertions of Pauli matrices in
the standard definition of the Wigner function~\cite{RaimondSI}. For instance, we can project
onto the $\ket{\uparrow}$ component only, by considering:
\begin{equation}
W^{(k)}_{\frac{1+\sigma^z}{2}}(X)=
 \int \!\!\! \frac{\mathrm{d^2}\lambda}{\pi^2}\;
e^{X(\bar{\lambda}-\lambda)} 
\big<GS\big| e^{\lambda \akd-\bar{\lambda}\ak}
\frac{1+\sigma^z}{2}
\big|GS\big>.
\label{SupWSz}
\end{equation}
This Wigner distribution can be addressed from our two-polaron
variational state~(\ref{variationalgs}).
In the case $\Delta\ll\omega_c$, we find $p_2\propto\Delta/\omega_c\ll p_1$, and the wavefunction
is normalized by taking simply $p_1\simeq1/\sqrt{2}$.
We thus need to compute:
\begin{equation}
W^{(k)}_{\frac{1+\sigma^z}{2}}(X)=
 \frac{1}{2} \int \!\!\! \frac{\mathrm{d^2}\lambda}{\pi^2}\;
e^{X(\bar{\lambda}-\lambda)} 
\left[ 
\big<f_q^\mathrm{pol.}\big| e^{\lambda \akd-\bar{\lambda}\ak}
\big|f_q^\mathrm{pol.}\big>
+ 2p_2
\big<f_q^\mathrm{pol.}\big| e^{\lambda \akd-\bar{\lambda}\ak}
\big|f_q^\mathrm{anti.}\big>
+ p_2^2
\big<f_q^\mathrm{anti.}\big| e^{\lambda \akd-\bar{\lambda}\ak}
\big|f_q^\mathrm{anti.}\big>
\right].
\label{SupWSzIntegral}
\end{equation} 
Using usual coherent state algebra, we readily obtain the required overlaps:
\begin{eqnarray}
\big<f_q^\mathrm{pol.}\big| e^{\lambda \akd-\bar{\lambda}\ak}
\big|f_q^\mathrm{pol.}\big> &=& 
e^{(\lambda-\bar\lambda) f^\mathrm{pol.}_k}
e^{-\lambda\bar\lambda/2}\\
\big<f_q^\mathrm{pol.}\big| e^{\lambda \akd-\bar{\lambda}\ak}
\big|f_q^\mathrm{anti.}\big> &=& e^{-\frac{1}{2}\sum_q
(f_q^\mathrm{pol.}-f_g^\mathrm{anti.})^2}
e^{\lambda f^\mathrm{pol.}_k- \bar\lambda f^\mathrm{anti.}_k} 
e^{-\lambda\bar\lambda/2}\\
\big<f_q^\mathrm{anti.}\big| e^{\lambda \akd-\bar{\lambda}\ak}
\big|f_q^\mathrm{anti.}\big> &=& 
e^{(\lambda-\bar\lambda) f^\mathrm{anti.}_k}
e^{-\lambda\bar\lambda/2}
\end{eqnarray}
and performing the Gaussian integral in Eq.~(\ref{SupWSzIntegral}) yields
\begin{equation}
W^{(k)}_{\frac{1+\sigma^z}{2}}(X)=
\frac{1}{\pi} e^{-2(X-f_k^\mathrm{pol.})^2}
+ \frac{2 p_2}{\pi}
e^{-\frac{1}{2}\sum_{q\neq k} (f^{\mathrm{pol.}}_{q}-f^{\mathrm{anti.}}_{q})^2}
 e^{-2\big(X-\frac{f^{\mathrm{pol.}}_{k}+f^{\mathrm{anti.}}_{k}}{2}\big)^2}
+\frac{p_2^2}{\pi} e^{-2(X-f_k^\mathrm{anti.})^2}
.
\label{SupWSzFinal}
\end{equation}
Note that the second term in Eq.~(\ref{SupWSzFinal}) provides only a small
correction to the first contribution, a feature which follows from
(i) $p_2\ll1$ and (ii) the overlap appearing in this second contribution
can be approximated as
$e^{-\frac{1}{2}\sum_{q\neq k} (f^{\mathrm{pol.}}_{q}-f^{\mathrm{anti.}}_{q})^2} 
\simeq \big<f_q^\mathrm{pol.}\big| f_q^\mathrm{anti.}\big>
\simeq e^{-2\sum_{q} (f^{\mathrm{pol.}}_{q})^2}\simeq \Delta_R/\Delta \ll 1$
(for $\alpha\gtrsim0.5$) because the antipolaron is equal and
opposite to the polaron at low energy. Similarly, the third term in
Eq.~(\ref{SupWSzFinal}), which would peak at the antipolaron displacement,
is of order $p_2^2\ll1$, and so also provides a tiny contribution.
Thus, the $\ket{\uparrow}$-projected Wigner function is dominated by the purely polaronic
contribution, as we indeed demonstrate by the impressive agreement with the numerically 
exact NRG computation of $W^{(k)}_{\frac{1+\sigma^z}{2}}(X)$ in the left panel of
Fig.~\ref{WignerCompare}.

In order to highlight the emergence of antipolaronic contributions in the
wavefunction, we now insert the off-diagonal $\sigma^+$ Pauli matrix,
which leads to the equivalent expression as defined in Eq.~(7) of the
main text: 
\begin{equation}
W^{(k)}_{\sigma^+}(X)=
 \int \!\!\! \frac{\mathrm{d^2}\lambda}{\pi^2}\;
e^{X(\bar{\lambda}-\lambda)} 
\big<GS\big| e^{\lambda \akd-\bar{\lambda}\ak}
\sigma^+
\big|GS\big>.
\end{equation}
We thus need to compute
\begin{eqnarray}
\label{SupWSplus}
\nonumber
W^{(k)}_{\sigma^+}(X)&=&
 \frac{1}{2} \int \!\!\! \frac{\mathrm{d^2}\lambda}{\pi^2}\;
e^{X(\bar{\lambda}-\lambda)} 
\Big[ 
\big<-f_q^\mathrm{pol.}\big| e^{\lambda \akd-\bar{\lambda}\ak}
\big|f_q^\mathrm{pol.}\big>
+ p_2
\big<-f_q^\mathrm{pol.}\big| e^{\lambda \akd-\bar{\lambda}\ak}
\big|f_q^\mathrm{anti.}\big>
+ p_2
\big<-f_q^\mathrm{anti.}\big| e^{\lambda \akd-\bar{\lambda}\ak}
\big|f_q^\mathrm{pol.}\big>
\\
&& + p_2^2
\big<-f_q^\mathrm{anti.}\big| e^{\lambda \akd-\bar{\lambda}\ak}
\big|f_q^\mathrm{anti.}\big>
\Big].
\end{eqnarray}
A computation similar to performed above leads to the final result:
\begin{eqnarray}
\label{SupWSplusFinal}
\nonumber
W_{\sigma^+}^{(k)}(X) &=& \frac{1}{\pi} e^{-2\sum_{q\neq k} (f_q^\mathrm{pol.})^2}
e^{-2X^2}
+\frac{p_2}{\pi}
e^{-\frac{1}{2}\sum_{q\neq k} (f^{\mathrm{pol.}}_{q}+f^{\mathrm{anti.}}_{q})^2}
\Big[
e^{-2\big(X-\frac{f^{\mathrm{pol.}}_{k}-f^{\mathrm{anti.}}_{k}}{2}\big)^2}
+ e^{-2\big(X+\frac{f^{\mathrm{pol.}}_{k}-f^{\mathrm{anti.}}_{k}}{2}\big)^2}\Big]
\\ 
&& +\frac{p_2^2}{\pi} e^{-2\sum_{q\neq k} (f_q^\mathrm{anti.})^2}
e^{-2X^2}\;.
\end{eqnarray}
The above expression shows important differences from the $\ket{\uparrow}$-projected Wigner
distribution of Eq.~(\ref{SupWSzFinal}). Indeed, the first term associated with the
purely polaronic response is now of order $e^{-2\sum_{q\neq k}
(f_q^\mathrm{pol.})^2}\simeq \big<f_q^\mathrm{pol.}\big| -f_q^\mathrm{pol.}\big>
\simeq \Delta_R/\Delta \ll 1$, and so is subdominant to the second
contribution (with mixed polaron-antipolaron origin) of order $p_2\propto \Delta/\omega_c$
(Note that the overlap $e^{-\frac{1}{2}\sum_{q\neq k}
(f^{\mathrm{pol.}}_{q}+f^{\mathrm{anti.}}_{q})^2}$ appearing in the second term
is of order 1). The third term, of order $p_2^2 \Delta_R/\Delta$ is even more smaller.
Thus, the off-diagonal Wigner function $W^{(k)}_{\sigma^+}(X)$ can be used to
highlight the emergence of antipolarons in the many-body ground state
wavefunction of the continuous spin-boson model, as was also discussed in the
main text.


\begin{thebibliography}{99}
\frenchspacing

\bibitem{Raimond} Raimond, J. M. \& Haroche, S.
{\it Understanding the Quantum} (Oxford Graduate Series, 2006).

\bibitem{nielsen}
Nielsen, A. M.  \& Chuang, I. L.  {\it Quantum Computation and Quantum 
Information} (Cambridge University Press, New York, 2007).

\bibitem{LambertNP2013}
Lambert, N., Chen, Y.-N., Cheng, Y.-C., Li, C.-J., Chen, G.-Y. \& Nori, F.
Quantum biology. {\it Nature Physics} {\bf 9}, 10 (2013).

\bibitem{scholes2011lessons}
Scholes, G., Fleming, G., Olaya-Castro, A. \& van Grondelle, R.
{\it Nature Chemistry}{ \bf 3}, 763 (2011).

\bibitem{Leggett} Leggett, A. J., Chakravarty, S., Dorsey, A. T.,
Fisher, M. P. A., Garg, A. \& Zwerger W. Dynamics of the dissipative two-state
system. {\it Rev. Mod. Phys.} {\bf 59}, 1 (1987).

\bibitem{Weiss} Weiss, U. {\it Quantum Dissipative Systems} (World Scientific, 1993).

\bibitem{BreuerPetruccione}
Breuer, H.-P. \& Petruccione, F. {\it The Theory of Open Quantum Systems,} (Oxford 
University Press, 2010).

\bibitem{NitzanBook}
Nitzan, A. {\it Chemical Dynamics in Condensed Phases: Relaxation, 
Transfer and Reactions in Condensed Molecular Systems} 
(Oxford University Press, 2006).

\bibitem{Jennings09}
Jennings, D., Dragan, A., Barrett, S. D., Bartlett, S. D. \& Rudolph, T. 
Quantum computation via measurements on the low-temperature state of a many-body system.
{\it Phys. Rev. A} \textbf{80}, 032328 (2009).

\bibitem{RaussendorfBriegel2001}
Raussendorf, R. \& Briegel, H. J. A one-way quantum computer. {\it Phys.
Rev. Lett.} \textbf{86}, 5188 (2001).

\bibitem{NRG-RMP08}
Bulla, R., Costi, T. A. \& Pruschke, T. Numerical renormalization group method for quantum 
impurity systems. {\it Rev. Mod. Phys.} \textbf{80}, 395 (2008).

\bibitem{Makri95}
Makri, N. Numerical path integral techniques for long time dynamics of quantum
dissipative systems. {\it J. Math. Phys.} \textbf{36}, 2430 (1995).

\bibitem{WangThoss08}
Wang, H. \& Thoss, M. From coherent motion to localization: dynamics of the spin-boson model at zero temperature. 
{\it New J. Phys.} \textbf{10}, 115005 (2008).

\bibitem{NalbachThorwart10}
Nalbach, P. \& Thorwart, M. Ultraslow quantum dynamics in a sub-ohmic heat bath. 
{\it Phys. Rev. B} \textbf{81}, 054308 (2010).

\bibitem{WinterBulla09}
Winter, A., Rieger, H., Vojta, M. \& Bulla, R. Quantum phase transition in the sub-ohmic spin-boson model: 
Quantum Monte-Carlo study with a continuous imaginary time cluster algorithm.
{\it Phys. Rev. Lett.} \textbf{102}, 030601 (2009).

\bibitem{AlvermannFehske09}
Alvermann, A. \& Fehske, H. Sparse polynomial space approach to dissipative quantum systems: 
Application to the sub-ohmic spin-boson model. {\it Phys. Rev. Lett.} \textbf{102}, 150601 (2009).

\bibitem{PriorPlenio_EffSim10}
Prior, J., Chin, A. W., Huelga, S. F. \& Plenio, M. B. Efficient simulation of strong system-environment 
interactions. {\it Phys. Rev. Lett.} \textbf{105}, 050404 (2010).

\bibitem{Florens_DissipSpinDyn11}
Florens, S., Freyn, A., Venturelli, D. \& Narayanan, R. Dissipative spin dynamics near a quantum 
critical point: Numerical renormalization group and Majorana diagrammatics. {\it
Phys. Rev. B} \textbf{84}, 155110 (2011).

\bibitem{Silbey} Silbey, R. \& Harris, R. Variational calculation of the dynamics of a two level system 
interacting with a bath. {\it J. Chem. Phys.} {\bf 80}, 2615 (1984).

\bibitem{HarrisSilbey85}
Harris, R. A. \& Silbey, R. Variational calculation of the tunneling system interacting with a heat bath. 
II. Dynamics of an asymmetric tunneling system. {\it J. Chem. Phys.} \textbf{83}, 1069 (1985).

\bibitem{Chin_SubohmSBM11}
Chin, A. W., Prior, J., Huelga, S. F. \& Plenio, M. B. Generalized polaron ansatz for the ground state 
of the sub-ohmic spin-boson model: An analytic theory of the localization
transition. {\it Phys. Rev. Lett.} \textbf{107}, 160601 (2011).

\bibitem{Nazir} Nazir, A., McCutcheon, D. P. S. \& Chin, A. W. 
Ground state and dynamics of the biased dissipative two-state system: Beyond
variational polaron theory.  {\it Phys. Rev. B} {\bf 85}, 224301 (2012).

\bibitem{Demler} Agarwal, K., Martin, I., Lukin, M. D. \& Demler, E.
Polaronic model of Two Level Systems in amorphous solids. {\it Preprint,
arxiv:1212.3299}.

\bibitem{Braak} Braak, D. Integrability of the Rabi model. {\it Phys. Rev.  Lett.} 
{\bf 107}, 100401 (2011).

\bibitem{hwang} Hwang, M.-Y. \& Choi, M.-S. Variational study of a two-level
system coupled to a harmonic oscillator in an ultrastrong-coupling regime.
{\it Phys. Rev. A.} {\bf 82}, 025802 (2010).

\bibitem{stolze} Stolze, J. \& M{\"u}ller, L. Quality of variational ground
states for a two-state system coupled to phonons. {\it Phys. Rev. B.} {\bf 42}, 
6704 (1990).

\bibitem{Bulla} Bulla, R., Tong, N.-H. \& Vojta, M. Numerical renormalization
group for bosonic systems and application to the sub-ohmic spin-boson model.
{\it Phys. Rev. Lett.} {\bf 91}, 170601 (2003).

\bibitem{ReviewTomography} 
Lvovsky, A. I. \& Raymer, M. G. Continuous-variable optical quantum-state tomography. 
{\it Rev. Mod. Phys.} \textbf{81}, 299 (2009).

\bibitem{Hofheinz} Hofheinz, M., Wang, H., Ansmann, M., Bialczak, R. C., Lucero, E.,
Neeley, M., O'Connell, A. D., Sank, D., Wenner, J., Martinis, J. M. \& Cleland, 
A. N. Synthesizing arbitrary quantum states in a superconducting resonator.
{\it Nature} {\bf 459}, 546 (2009). 

\bibitem{Eichler} Eichler, C., Bozyigit, D., Lang, C., Steffen, L., Fink, J. \& 
Wallraff, A. Experimental state tomography of itinerant single microwave photons.
{\it Phys. Rev. Lett.} {\bf 106}, 220503 (2011).

\bibitem{LeHur} Le Hur, K. Kondo resonance of a microwave photon. {\it Phys.
Rev. B} {\bf 85}, 140506 (2012).
 
\bibitem{BallesterSolano}
 Ballester, D., Romero, G., Garc\'{\i}a-Ripoll, J. J., Deppe, F. \& Solano, E.
Quantum simulation of the ultrastrong-coupling dynamics in circuit quantum
electrodynamics. {\it Phys. Rev. X} {\bf 2}, 021007 (2012).

\bibitem{Goldstein} Goldstein, M., Devoret, M. H., Houzet, M. \& Glazman, L. I. 
Inelastic microwave photon scattering off a quantum impurity in a
Josephson-junction array. {\it Phys. Rev. Lett.} {\bf 110}, 017002 (2013).


\end{thebibliography}

\begin{thebibliography}{99}
\frenchspacing

\bibitem{LeggettSI} Leggett, A. J., Chakravarty, S., Dorsey, A. T.,
Fisher, M. P. A., Garg, A. \& Zwerger W. Dynamics of the dissipative two-state
system. {\it Rev. Mod. Phys.} {\bf 59}, 1 (1987).

\bibitem{WeissSI} Weiss, U. {\it Quantum Dissipative Systems} (World Scientific, 1993).

\bibitem{SilbeySI} Silbey, R. \& Harris, R. Variational calculation of the dynamics of a two level system 
interacting with a bath. {\it J. Chem. Phys.} {\bf 80}, 2615 (1984).

\bibitem{Chin_SubohmSBM11SI}
Chin, A. W., Prior, J., Huelga, S. F. \& Plenio, M. B. Generalized polaron ansatz for the ground state 
of the sub-ohmic spin-boson model: An analytic theory of the localization
transition. {\it Phys. Rev. Lett.} \textbf{107}, 160601 (2011).

\bibitem{NazirSI} Nazir, A., McCutcheon, D. P. S. \& Chin, A. W. 
Ground state and dynamics of the biased dissipative two-state system: Beyond
variational polaron theory.  {\it Phys. Rev. B} {\bf 85}, 224301 (2012).

\bibitem{BraakSI} Braak, D. Integrability of the Rabi model. {\it Phys. Rev.  Lett.} 
{\bf 107}, 100401 (2011).

\bibitem{RaimondSI} Raimond, J. M. \& Haroche, S.
{\it Understanding the Quantum} (Oxford Graduate Series, 2006).

\end{thebibliography}
\end{document}